%% file: main.tex
\pdfoutput=1
\documentclass[12pt,a4paper]{article}

\def\BdKsMuMu     {\decay{\Bd}{ \KS \mup\mun}}
\def\BdKzMuMu     {\decay{\Bd}{ \Kz \mup\mun}}
\def\BdJpsiKs     {\decay{\Bd}{ \jpsi \KS}}
\def\BuJpsiK       {\decay{\Bp}{ \jpsi \Kp}}
\def\BuKMuMu     {\decay{\Bp}{ \Kp \mup\mun}}
\def\KsMuMu     {\ensuremath{ \KS\mup\mun}}
\def\KMuMu    {\ensuremath { \Kp\mup\mun}}
\def\bukst     {\ensuremath{ \Kstarp\mup\mun}}
\def\KstMuMu     {\ensuremath{ \Kstarz\mup\mun}}

\def\BuKstKsPiMuMu     {\decay{\Bp}{ (K^{*+} \to \KS \pip) \mup\mun}}
\def\BuKstMuMu     {\decay{\Bp}{ K^{*+} \mup\mun}}

\def\BdKstMuMu     {\decay{\Bd}{ \Kstarz \mup\mun}}
\def\BdKPiMuMu {\decay{\Bd}{(\Kstarz \to \Kp \pim) \mup\mun}}
\def\BdJpsiKst    {\decay{\Bd}{ \jpsi K^{*0} }}
\def\BKMuMu    {\decay{\B}{ K \mup\mun}}
\def\BKstMuMu {\decay{\B}{ K^{*} \mup \mun}}

\def\AllModes    {\decay{\B}{ K^{(*)} \mup\mun}}
\def\CharModes    {\decay{\Bu}{ K^{(*)+} \mup\mun}}
\def\NeutModes    {\decay{\Bd}{ K^{(*)0} \mup\mun}}

\def\qq {\ensuremath{q^{2}}\xspace}
\def\AI {\ensuremath{A_{\rm I}}\xspace}

\def\JpsiModes {\decay{\B}{ \jpsi K^{(*)}  }}

\usepackage{lineno}  
\usepackage{graphicx}  
\usepackage{xspace} 
\usepackage{color}
\usepackage{colortbl}
\usepackage{amsmath} 
\usepackage{ifthen} 
\usepackage{subfigure}
\newboolean{pdflatex}
\setboolean{pdflatex}{true} 
\newboolean{articletitles}
\setboolean{articletitles}{true} 

\newboolean{uprightparticles}
\setboolean{uprightparticles}{false} 

\usepackage{amssymb}
\usepackage{amsfonts}
\usepackage{upgreek} 
\usepackage{ctable}

\usepackage{hyperref}    
\usepackage[all]{hypcap} 

\input{lhcb-symbols-def} 

\usepackage{mciteplus}

\begin{document}

\renewcommand{\thefootnote}{\fnsymbol{footnote}}
\setcounter{footnote}{1}

\input{title-LHCb-PAPER}


\renewcommand{\thefootnote}{\arabic{footnote}}
\setcounter{footnote}{0}


\pagestyle{plain} 
\setcounter{page}{1}
\pagenumbering{arabic}

 \linenumbers

\input{introduction}

\input{selection}

\input{massfits}

\input{normalisation}

\input{systematics}

\input{results}

\input{acknowledgements}

\bibliographystyle{LHCb}
\bibliography{main}

\end{document}

%% file: lhcb-symbols-def.tex
\def\lhcb {\mbox{LHCb}\xspace}
\def\ux85 {\mbox{UX85}\xspace}

\def\babar  {\mbox{BaBar}\xspace}

\ifthenelse{\boolean{uprightparticles}}%
{

 \def\Pmu         {\ensuremath{\upmu}\xspace}

 \def\Ppi         {\ensuremath{\uppi}\xspace}

 \def\Ppsi        {\ensuremath{\uppsi}\xspace}

 \def\PDelta      {\ensuremath{\Delta}\xspace}                 
 \def\PXi      {\ensuremath{\Xi}\xspace}                 
 \def\PLambda      {\ensuremath{\Lambda}\xspace}                 
 \def\PSigma      {\ensuremath{\Sigma}\xspace}                 
 \def\POmega      {\ensuremath{\Omega}\xspace}                 
 \def\PUpsilon      {\ensuremath{\Upsilon}\xspace}                 
 
 \def\PB      {\ensuremath{\mathrm{B}}\xspace}                 
                  
 \def\PD      {\ensuremath{\mathrm{D}}\xspace}

 \def\PJ      {\ensuremath{\mathrm{J}}\xspace}                 
 \def\PK      {\ensuremath{\mathrm{K}}\xspace}

 \def\Pb      {\ensuremath{\mathrm{b}}\xspace}                 
 \def\Pc      {\ensuremath{\mathrm{c}}\xspace}

 \def\Pi      {\ensuremath{\mathrm{i}}\xspace}

 \def\Pp      {\ensuremath{\mathrm{p}}\xspace}

 \def\Ps      {\ensuremath{\mathrm{s}}\xspace}

}
{

 \def\Pmu         {\ensuremath{\mu}\xspace}

 \def\Ppi         {\ensuremath{\pi}\xspace}

 \def\Ppsi        {\ensuremath{\psi}\xspace}                 
                  
 \mathchardef\PDelta="7101
 \mathchardef\PXi="7104
 \mathchardef\PLambda="7103
 \mathchardef\PSigma="7106
 \mathchardef\POmega="710A
 \mathchardef\PUpsilon="7107
                  
 \def\PB      {\ensuremath{B}\xspace}                 
                  
 \def\PD      {\ensuremath{D}\xspace}

 \def\PJ      {\ensuremath{J}\xspace}                 
 \def\PK      {\ensuremath{K}\xspace}

 \def\Pb      {\ensuremath{b}\xspace}                 
 \def\Pc      {\ensuremath{c}\xspace}

 \def\Pi      {\ensuremath{i}\xspace}

 \def\Pp      {\ensuremath{p}\xspace}

 \def\Ps      {\ensuremath{s}\xspace}

}

\def\mup        {\ensuremath{\Pmu^+}\xspace}
\def\mun        {\ensuremath{\Pmu^-}\xspace} 
\def\mumu       {\ensuremath{\Pmu^+\Pmu^-}\xspace}

\def\squark    {\ensuremath{\Ps}\xspace}

\def\cquark    {\ensuremath{\Pc}\xspace}

\def\bquark    {\ensuremath{\Pb}\xspace}

\def\pion  {\ensuremath{\Ppi}\xspace}
\def\piz   {\ensuremath{\pion^0}\xspace}

\def\pip   {\ensuremath{\pion^+}\xspace}
\def\pim   {\ensuremath{\pion^-}\xspace}

\def\kaon  {\ensuremath{\PK}\xspace}
  \def\Kbar  {\kern 0.2em\overline{\kern -0.2em \PK}{}\xspace}

\def\Kz    {\ensuremath{\kaon^0}\xspace}
\def\Kzb   {\ensuremath{\Kbar^0}\xspace}
\def\KzKzb {\ensuremath{\Kz \kern -0.16em \Kzb}\xspace}
\def\Kp    {\ensuremath{\kaon^+}\xspace}
\def\Km    {\ensuremath{\kaon^-}\xspace}

\def\KpKm  {\ensuremath{\Kp \kern -0.16em \Km}\xspace}
\def\KS    {\ensuremath{\kaon^0_{\rm\scriptscriptstyle S}}\xspace} 
\def\KL    {\ensuremath{\kaon^0_{\rm\scriptscriptstyle L}}\xspace} 
\def\Kstarz  {\ensuremath{\kaon^{*0}}\xspace}

\def\Kstar   {\ensuremath{\kaon^*}\xspace}

\def\Kstarp  {\ensuremath{\kaon^{*+}}\xspace}

  \def\Dbar    {\kern 0.2em\overline{\kern -0.2em \PD}{}\xspace}
\def\D       {\ensuremath{\PD}\xspace}

\def\Dz      {\ensuremath{\D^0}\xspace}
\def\Dzb     {\ensuremath{\Dbar^0}\xspace}
\def\DzDzb   {\ensuremath{\Dz {\kern -0.16em \Dzb}}\xspace}
\def\Dp      {\ensuremath{\D^+}\xspace}
\def\Dm      {\ensuremath{\D^-}\xspace}

\def\DpDm    {\ensuremath{\Dp {\kern -0.16em \Dm}}\xspace}

\def\Dstarp  {\ensuremath{\D^{*+}}\xspace}

\def\B       {\ensuremath{\PB}\xspace}
  \def\Bbar    {\kern 0.18em\overline{\kern -0.18em \PB}{}\xspace}

\def\Bz      {\ensuremath{\B^0}\xspace}

\def\Bu      {\ensuremath{\B^+}\xspace}

\def\Bp      {\ensuremath{\Bu}\xspace}

\def\Bd      {\ensuremath{\B^0}\xspace}
\def\Bs      {\ensuremath{\B^0_\squark}\xspace}

\def\jpsi     {\ensuremath{{\PJ\mskip -3mu/\mskip -2mu\Ppsi\mskip 2mu}}\xspace}
\def\psitwos  {\ensuremath{\Ppsi{(2S)}}\xspace}

  \def\Y#1S{\ensuremath{\PUpsilon{(#1S)}}\xspace}

\def\proton      {\ensuremath{\Pp}\xspace}

\def\L {\ensuremath{\PLambda}\xspace}
\def\Lbar {\ensuremath{\kern 0.1em\overline{\kern -0.1em\PLambda}}\xspace}

\def\Lb      {\ensuremath{\L^0_\bquark}\xspace}

\newcommand{\decay}[2]{\ensuremath{#1\!\to #2}\xspace}         

\def\to                 {\ensuremath{\rightarrow}\xspace}

\def\qsq       {\ensuremath{q^2}\xspace}

\def\CP                {\ensuremath{C\!P}\xspace}

\def\AT#1     {\ensuremath{A_{\mathrm{T}}^{#1}}\xspace}           

\def\C#1      {\ensuremath{\mathcal{C}_{#1}}\xspace}                       
\def\Cp#1     {\ensuremath{\mathcal{C}_{#1}^{'}}\xspace}                    
\def\Ceff#1   {\ensuremath{\mathcal{C}_{#1}^{\mathrm{(eff)}}}\xspace}        
\def\Cpeff#1  {\ensuremath{\mathcal{C}_{#1}^{'\mathrm{(eff)}}}\xspace}       
\def\Ope#1    {\ensuremath{\mathcal{O}_{#1}}\xspace}                       
\def\Opep#1   {\ensuremath{\mathcal{O}_{#1}^{'}}\xspace}                    

\newcommand{\tev}{\ensuremath{\mathrm{\,Te\kern -0.1em V}}\xspace}
\newcommand{\gev}{\ensuremath{\mathrm{\,Ge\kern -0.1em V}}\xspace}
\newcommand{\mev}{\ensuremath{\mathrm{\,Me\kern -0.1em V}}\xspace}
\newcommand{\kev}{\ensuremath{\mathrm{\,ke\kern -0.1em V}}\xspace}
\newcommand{\ev}{\ensuremath{\mathrm{\,e\kern -0.1em V}}\xspace}
\newcommand{\gevc}{\ensuremath{{\mathrm{\,Ge\kern -0.1em V\!/}c}}\xspace}
\newcommand{\mevc}{\ensuremath{{\mathrm{\,Me\kern -0.1em V\!/}c}}\xspace}
\newcommand{\gevcc}{\ensuremath{{\mathrm{\,Ge\kern -0.1em V\!/}c^2}}\xspace}
\newcommand{\gevgevcccc}{\ensuremath{{\mathrm{\,Ge\kern -0.1em V^2\!/}c^4}}\xspace}
\newcommand{\mevcc}{\ensuremath{{\mathrm{\,Me\kern -0.1em V\!/}c^2}}\xspace}

\def\mum  {\ensuremath{\,\upmu\rm m}\xspace}

\def\invfb   {\ensuremath{\mbox{\,fb}^{-1}}\xspace}

\def\ps   {\ensuremath{{\rm \,ps}}\xspace}

\def\gsim{{~\raise.15em\hbox{$>$}\kern-.85em
          \lower.35em\hbox{$\sim$}~}\xspace}
\def\lsim{{~\raise.15em\hbox{$<$}\kern-.85em
          \lower.35em\hbox{$\sim$}~}\xspace}

\def\pt         {\mbox{$p_{\rm T}$}\xspace}

\def\dllkpi     {\ensuremath{\mathrm{DLL}_{\kaon\pion}}\xspace}
\def\dllppi     {\ensuremath{\mathrm{DLL}_{\proton\pion}}\xspace}

\def\evtgen     {\mbox{\textsc{EvtGen}}\xspace}
\def\pythia     {\mbox{\textsc{Pythia}}\xspace}

\def\geant      {\mbox{\textsc{Geant4}}\xspace}

\def\photos     {\mbox{\textsc{Photos}}\xspace}

\def\tell1  {TELL1\xspace}
\def\ukl1   {UKL1\xspace}



%% file: title-LHCb-PAPER.tex

\begin{titlepage}
\pagenumbering{roman}

\vspace*{-1.5cm}
\centerline{\large EUROPEAN ORGANIZATION FOR NUCLEAR RESEARCH (CERN)}
\vspace*{1.5cm}
\hspace*{-0.5cm}
\begin{tabular*}{\linewidth}{lc@{\extracolsep{\fill}}r}
\ifthenelse{\boolean{pdflatex}}
{\vspace*{-2.7cm}\mbox{\!\!\!\includegraphics[width=.14\textwidth]{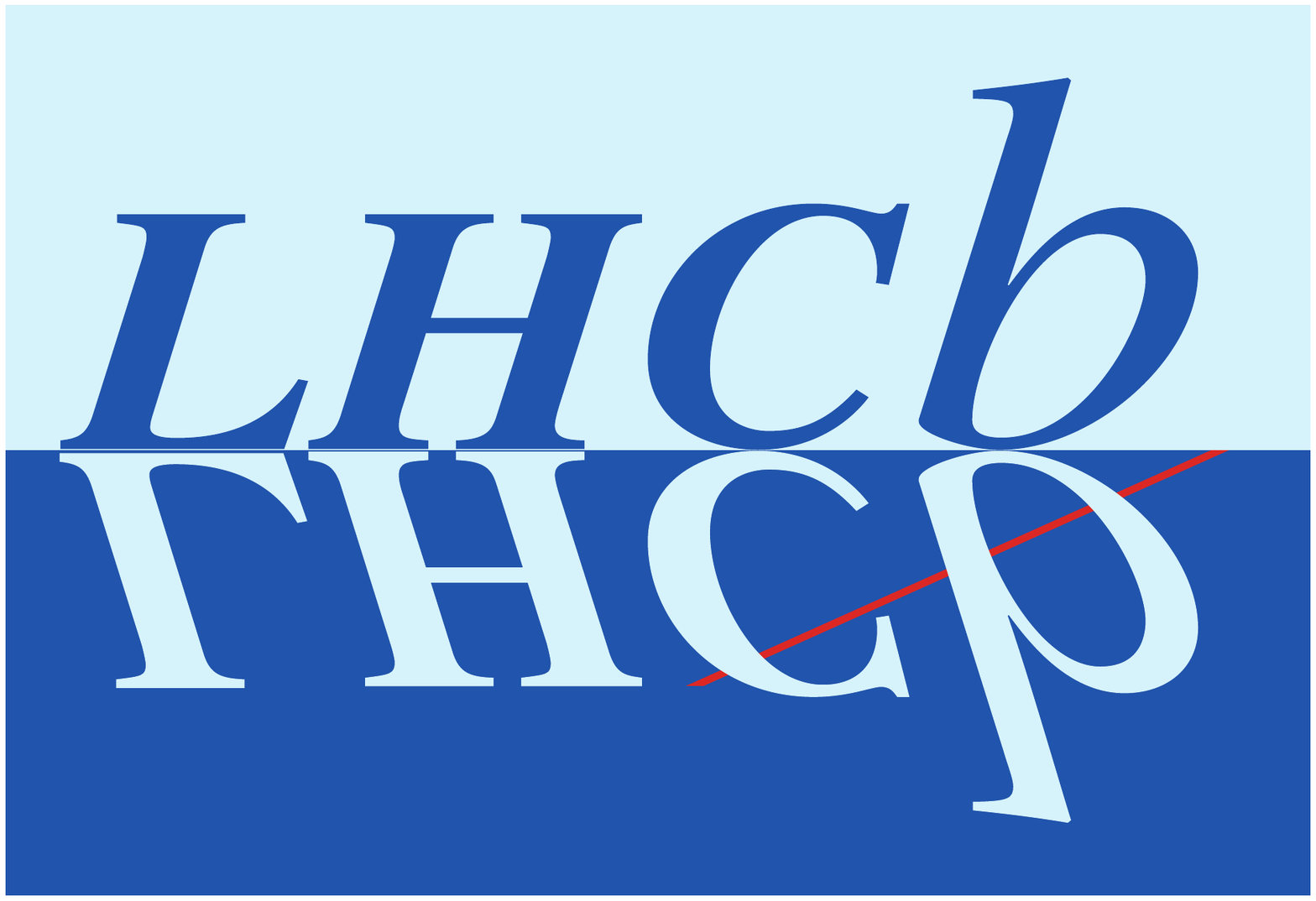}} & &}%
{\vspace*{-1.2cm}\mbox{\!\!\!\includegraphics[width=.12\textwidth]{figs/lhcb-logo.eps}} & &}%
\\
 & & CERN-PH-EP-2012-129 \\  
 & & LHCb-PAPER-2012-011 \\  
 & & May 15, 2012 \\ 
 & & \\
\end{tabular*}

\vspace*{3.5cm}

{\bf\boldmath\huge
\begin{center}
  Measurement of the isospin asymmetry in \AllModes decays
\end{center}
}

\vspace*{2.0cm}

\begin{center}
The LHCb collaboration\footnote{Authors are listed on the following pages.}
\end{center}

\vspace{\fill}

\begin{abstract}
\noindent
\input{abstract}
\end{abstract}

\vspace*{1.0cm}

\begin{center}
  Submitted to Journal of High Energy Physics
\end{center}

\vspace{\fill}

\end{titlepage}


\newpage
\setcounter{page}{2}
\mbox{~}
\newpage

\input{LHCb_authorlist.tex}

\cleardoublepage

%% file: abstract.tex
The isospin asymmetries of $B \to K^{(*)}\mu^+\mu^-$ decays and the
partial branching fractions of $B^0 \to K^0\mu^+\mu^-$ and $B^+ \to
K^{*+}\mu^+\mu^-$ are measured as a function of the di-muon mass
squared $q^2$ using an integrated luminosity of 1.0 fb$^{-1}$
collected with the LHCb detector. The $B \to K\mu^+\mu^-$ isospin
asymmetry integrated over $q^2$ is negative, deviating from zero with
over 4 $\sigma$ significance. The $B \to K^{*}\mu^+\mu^-$ decay
measurements are consistent with the Standard Model prediction of
negligible isospin asymmetry. The observation of the decay $B^0 \to
K^0_{\rm\scriptscriptstyle S}\mu^+\mu^-$ is reported with 5.7 $\sigma$
significance. Assuming that the branching fraction of $B^0 \to
K^0\mu^+\mu^-$ is twice that of $B^0 \to K^0_{\rm\scriptscriptstyle
S}\mu^+\mu^-$, the branching fractions of $B^0 \to K^0\mu^+\mu^-$ and
$B \to K^{*+}\mu^+\mu^-$ are found to be ($0.31^{+0.07}_{-0.06})
\times 10^{-6}$ and ($1.16\pm0.19) \times 10^{-6}$, respectively.

%% file: LHCb_authorlist.tex
\centerline{\large\bf LHCb collaboration}
\begin{flushleft}
\small
R.~Aaij$^{38}$, 
C.~Abellan~Beteta$^{33,n}$, 
A.~Adametz$^{11}$, 
B.~Adeva$^{34}$, 
M.~Adinolfi$^{43}$, 
C.~Adrover$^{6}$, 
A.~Affolder$^{49}$, 
Z.~Ajaltouni$^{5}$, 
J.~Albrecht$^{35}$, 
F.~Alessio$^{35}$
% ...

{\footnotesize \it
% ...
$ ^{5}$Clermont Universit\'{e}, Universit\'{e} Blaise Pascal, CNRS/IN2P3, LPC, Clermont-Ferrand, France\\
$ ^{6}$CPPM, Aix-Marseille Universit\'{e}, CNRS/IN2P3, Marseille, France\\
% ...
$ ^{11}$Physikalisches Institut, Ruprecht-Karls-Universit\"{a}t Heidelberg, Heidelberg, Germany\\
% ...
$ ^{33}$Universitat de Barcelona, Barcelona, Spain\\
$ ^{34}$Universidad de Santiago de Compostela, Santiago de Compostela, Spain\\
$ ^{35}$European Organization for Nuclear Research (CERN), Geneva, Switzerland\\
% ...
$ ^{38}$Nikhef National Institute for Subatomic Physics, Amsterdam, The Netherlands\\
% ...
$ ^{43}$H.H. Wills Physics Laboratory, University of Bristol, Bristol, United Kingdom\\
% ...
$ ^{49}$Oliver Lodge Laboratory, University of Liverpool, Liverpool, United Kingdom\\
% ...
$ ^{n}$LIFAELS, La Salle, Universitat Ramon Llull, Barcelona, Spain\\
% ...
}
\end{flushleft}

%% file: introduction.tex

\section{Introduction}
\label{sec:Introduction}

The flavour-changing neutral current decays \AllModes are forbidden at tree level in the Standard Model (SM). Such transitions must proceed via loop or box diagrams and are powerful probes of physics beyond the SM. Predictions for the branching fractions of these decays suffer from relatively large uncertainties due to form factor estimates. Theoretically clean observables can be constructed from ratios or asymmetries where the leading form factor uncertainties cancel. The \CP averaged isospin asymmetry (\AI) is such an observable. It is defined as

\begin{equation} 
\begin{split}
\AI = \frac{\Gamma(\NeutModes) -\Gamma(\CharModes)}{\Gamma(\NeutModes) +\Gamma(\CharModes)}\phantom{\AI}\phantom{,} \\ 
 = \frac{\mathcal{B}(\NeutModes) - \frac{\tau_0}{\tau_+}\mathcal{B}(\CharModes)}{\mathcal{B}(\NeutModes) + \frac{\tau_0}{\tau_+}\mathcal{B}(\CharModes)},
\end{split}
\label{eq:AI}
\end{equation}
where $\Gamma(\B\to f)$ and $\mathcal{B}(\B\to f)$ are the partial
width and branching fraction of the $\B\to f$ decay and
$\tau_0/\tau_+$ is the ratio of the lifetimes of the \Bz and \Bp
mesons.\footnote[1]{Charge conjugation is implied throughout this
  paper.} For \BKstMuMu, the SM prediction for \AI is around $-1\%$ in
the di-muon mass squared (\qq) region below the \jpsi resonance, apart
from the very low \qq region where it rises to $\mathcal{O}(10\%)$ as
\qq approaches zero~\cite{isospintheory}. There is no precise
prediction for \AI in the \BKMuMu case, but it is also expected to be
close to zero. The small isospin asymmetry predicted in the SM is due
to initial state radiation of the spectator quark, which is different
between the neutral and charged decays. Previously, \AI has been
measured to be significantly below zero in the \qq region below the
\jpsi resonance~\cite{babariso,belleiso}. In particular, the combined
\BKMuMu and \BKstMuMu isospin asymmetries measured by the \babar
experiment were 3.9 $\sigma$ below zero. For \BKstMuMu, \AI is
expected to be consistent with the $\B\to\Kstarz\gamma$ measurement of
$5\pm3\%$~\cite{HFAG} as \qq approaches zero. No such constraint is
present for \BKMuMu.

The isospin asymmetries are determined by measuring the differential
branching fractions of \mbox{\BuKMuMu}, \mbox{\BdKsMuMu},
\mbox{\BdKPiMuMu} and \mbox{\BuKstKsPiMuMu}; the decays involving a
\KL or \piz are not considered. The \KS meson is reconstructed via the
$\KS\to\pip\pim$ decay mode. The signal selections (Section~\ref{sec:Selection}) are optimised to
provide the lowest overall uncertainty on the isospin
asymmetries; this leads to a very tight selection for the
\mbox{\BuKMuMu} and \mbox{\BdKPiMuMu} channels where signal yield is sacrificed
to achieve overall uniformity with the \mbox{\BdKsMuMu} and
\mbox{\BuKstKsPiMuMu} channels, respectively.  In order to convert a
signal yield into a branching fraction, the four signal channels are
normalised to the corresponding \JpsiModes channels (Section~\ref{sec:Normalisation}). The relative
normalisation in each \qq bin is performed by calculating the relative
efficiency between the signal and normalisation channels using
simulated events. The normalisation of \BdKzMuMu assumes that
$\mathcal{B}(\BdKzMuMu) = 2\mathcal{B}(\BdKsMuMu)$.  Finally, \AI is
determined by simultaneously fitting the $K^{(*)}\mu^{+}\mu^{-}$ mass
distributions for all signal channels. Confidence intervals are
estimated for \AI using a profile likelihood method (Section~\ref{sec:Results}). Systematic
uncertainties are included in the fit using Gaussian constraints.

\section{Experimental setup}

The measurements described in this paper are performed with 1.0\invfb
of $pp$ collision data collected with the LHCb detector at the CERN
LHC during 2011. The \lhcb detector~\cite{Alves:2008zz} is a
single-arm forward spectrometer covering the pseudorapidity range
$2<\eta <5$, designed for the study of particles containing \bquark or
\cquark quarks. The detector includes a high precision tracking system
consisting of a silicon-strip vertex detector (VELO) surrounding the
$pp$ interaction region, a large-area silicon-strip detector~(TT)
located upstream of a dipole magnet with a bending power of about
$4{\rm\,Tm}$, and three stations of silicon-strip detectors and straw
drift-tubes placed downstream. The combined tracking system has a
momentum resolution $\Delta p/p$ that varies from 0.4\% at 5\gevc to
0.6\% at 100\gevc, and an impact parameter~(IP) resolution of 20\mum for
tracks with high transverse momentum. Charged hadrons are identified
using two ring-imaging Cherenkov (RICH) detectors. Photon, electron
and hadron candidates are identified by a calorimeter system
consisting of scintillating-pad and pre-shower detectors, an
electromagnetic calorimeter and a hadronic calorimeter. Muons are
identified by a muon system composed of alternating layers of iron and
multiwire proportional chambers.

The trigger consists of a hardware stage, based on information from
the calorimeter and muon systems, followed by a software stage which
applies a full event reconstruction.  For this analysis, candidate
events are first required to pass a hardware trigger which selects
muons with a transverse momentum, $\pt>1.48\gevc$ for one muon, and
0.56 and 0.48\gevc for two muons.  In the subsequent software
trigger~\cite{LHCb-PUB-2011-016}, at least one of the final state
particles is required to have both $\pt>0.8\gevc$ and IP
$>100\mum$ with respect to all of the primary proton-proton
interaction vertices in the event. Finally, the tracks of two or more
of the final state particles are required to form a vertex which is
significantly displaced from the primary vertices in the event.

For the simulation, $pp$ collisions are generated using
\pythia~6.4~\cite{Sjostrand:2006za} with a specific \lhcb
configuration~\cite{LHCb-PROC-2010-056}.  Decays of hadronic particles
are described by \evtgen~\cite{Lange:2001uf} in which final state
radiation is generated using \photos~\cite{Golonka:2005pn}. 
The \evtgen physics model used is based on Ref.~\cite{Ali:1999mm}. The
interaction of the generated particles with the detector and its
response are implemented using the \geant
toolkit~\cite{Allison:2006ve, *Agostinelli:2002hh} as described in
Ref.~\cite{LHCb-PROC-2011-006}.

%% file: selection.tex

\section{Event selection}
\label{sec:Selection}

Candidates are reconstructed with an initial cut-based selection,
which is designed to reduce the dataset to a manageable level.
Channels involving a \KS meson are referred to as \KS channels whereas
those with a \Kp meson are referred to as \Kp channels.  Only events
which are triggered independently of the \Kp candidate are
accepted. Therefore, apart from a small contribution from candidates
which are triggered by the \KS meson, the \KS and the \Kp channels are
triggered in a similar way. The initial selection places requirements
on the geometry, kinematics and particle identification (PID)
information of the signal candidates. Kaons are identified using
information from the RICH detectors, such as the difference in
log-likelihood (DLL) between the kaon and pion hypothesis, \dllkpi.
Kaon candidates are required to have \dllkpi $> 1$, which has a kaon
efficiency of $\sim85\%$ and a pion efficiency of $\sim10\%$. Muons
are identified using the amount of hits in the muon stations combined
with information from the calorimeter and RICH systems. The muon PID
efficiency is around 90\%.  Candidate \KS are required to have a
di-pion mass within 30 \mevcc of the nominal \KS mass and \Kstar
candidates are required to have an mass within 100\mevcc of the
nominal \Kstar mass.  At this stage, the \KS channels are split into
two categories depending on how the pions from the \KS decay are
reconstructed.  For decays where both pions have hits inside the VELO
and the downstream tracking detectors the \KS candidates are
classified as long (L).  If the daughter pions are reconstructed
without VELO hits (but still with TT hits upstream of the magnet) they
are classified as downstream (D) \KS candidates. Separate selections
are applied to the L and D categories in order to maximise the
sensitivity.  The selection criteria described in the next paragraph
refer to the \KS channels.  \\

After the initial selection, the L category has a much lower level of
background than the D category. For this reason simple cut-based
selections are applied to the former, whereas multivariate selections
are employed for the latter. Both \Bz and \Bp L selections require the
\KS decay time to be greater than 3\ps, and for the IP $\chi^2$ to be
greater than 10 when the IP of the \KS, with respect to the PV, is
forced to be zero. The \BdKzMuMu L selection requires that \mbox{\KS
  \pt $>$ 1\gevc} and \mbox{\B \pt $>$ 2\gevc}.  The \KS mass window
is also tightened to $\pm$20\mevcc. The \mbox{\BuKstMuMu} L selection
requires that the pion from the \Kstarp has an IP $\chi^{2} > 30$.
Multi-variate selections are applied to the D categories using a
boosted decision tree (BDT)~\cite{tmva} which uses geometrical and
kinematic information of the \B candidate and of its daughters. The
most discriminating variables according to the \Bz and \Bp BDTs are
the \KS \pt and the angle between the \B momentum and its line of
flight (from the primary vertex to the decay vertex).  The BDTs are
trained and tested on simulated events for the signal and data for the
background. The simulated events have been corrected to match the data
as described in Sect.~\ref{sec:Normalisation}. All the variables used
in the BDTs are well described in the simulation after correction. The
background sample used is 25\% of \B candidates which have
\mbox{$|m_{K^{(*)}\mu^{+}\mu^{-}} - m_{B}|$ $>$ 60\mevcc}, where
$m_{B}$ is obtained from fits to the appropriate \JpsiModes
normalisation channel.  These data are excluded from the analysis. The
selection based on the BDT output maximises the metric $S/\sqrt{S+B}$,
where $S$ and $B$ are the expected signal and background yields,
respectively.

The \Kp channels have, as far as possible, the same selection 
criteria as used to select the \KS channels. The cut-based selections applied 
to the L categories have the \KS specific variables (e.g. \KS decay time) 
removed and the remaining requirements are applied to the \Kp channels. 
The BDTs trained on the D categories contain variables which can be 
applied to both \KS and \Kp candidates and the BDTs trained on the \KS 
channels are simply applied to the corresponding \Kp channels. The \Kp channels 
are therefore also split into two different categories, one of which has the L 
selection applied, while the other one has the D selection applied. 
The overlap of events between these categories induces a correlation 
between the L and D categories for the \Kp channels. This correlation is accounted for in the fit to \AI. 

The final selection reduces the combinatorial background remaining after the initial selection by a factor of 5--20, 
while retaining 60--90\% of the signal, depending on the category and decay mode.
It is ineffective at reducing background from fully reconstructed \B decays, where one or 
more final state particles have been misidentified. Additional selection criteria are therefore applied. 
For the \KS channels, the $\L \to \proton \pim$ decay can be mistaken for a $\KS \to \pip \pim$ decay if the proton is misidentified as a pion. 
If one of the pion daughters from the \KS candidate has a \dllppi $>$ 10, the proton 
mass hypothesis is assigned to it. For the L(D) categories, if the $\proton\pim$ mass 
is within 10(15)\mevcc of the nominal \L mass the candidate is rejected.
This selection eliminates background from \decay{\Lb}{(\L \to \proton \pim) \mup \mun} 
which peaks above the \B mass.
 For the \BdKstMuMu decay, the same peaking background vetoes are used as in 
 Ref.~\cite{kstmumu}, which remove contaminations from $\Bs \to \phi \mup \mun$, \BdJpsiKst and \BdKstMuMu 
 decays where the kaon and pion are swapped. Finally, for the \BuKMuMu decay, 
 backgrounds from \BuJpsiK and $\B\to\psitwos\Kp$ are present, where the \Kp and \mup candidates are swapped.
 If a candidate has a $\Kp\mun$ track combination consistent with originating from a \jpsi or \psitwos resonance, the kaon is required to be 
 inside the acceptance of the muon system but to have insufficient hits in the muon stations to be classified as a muon. 
 These vetoes remove less than 1\% of the signal and reduce peaking backgrounds
 to a negligible level. \\  

\begin{figure}[t] 
 \centering 
\includegraphics[width=0.80\textwidth]{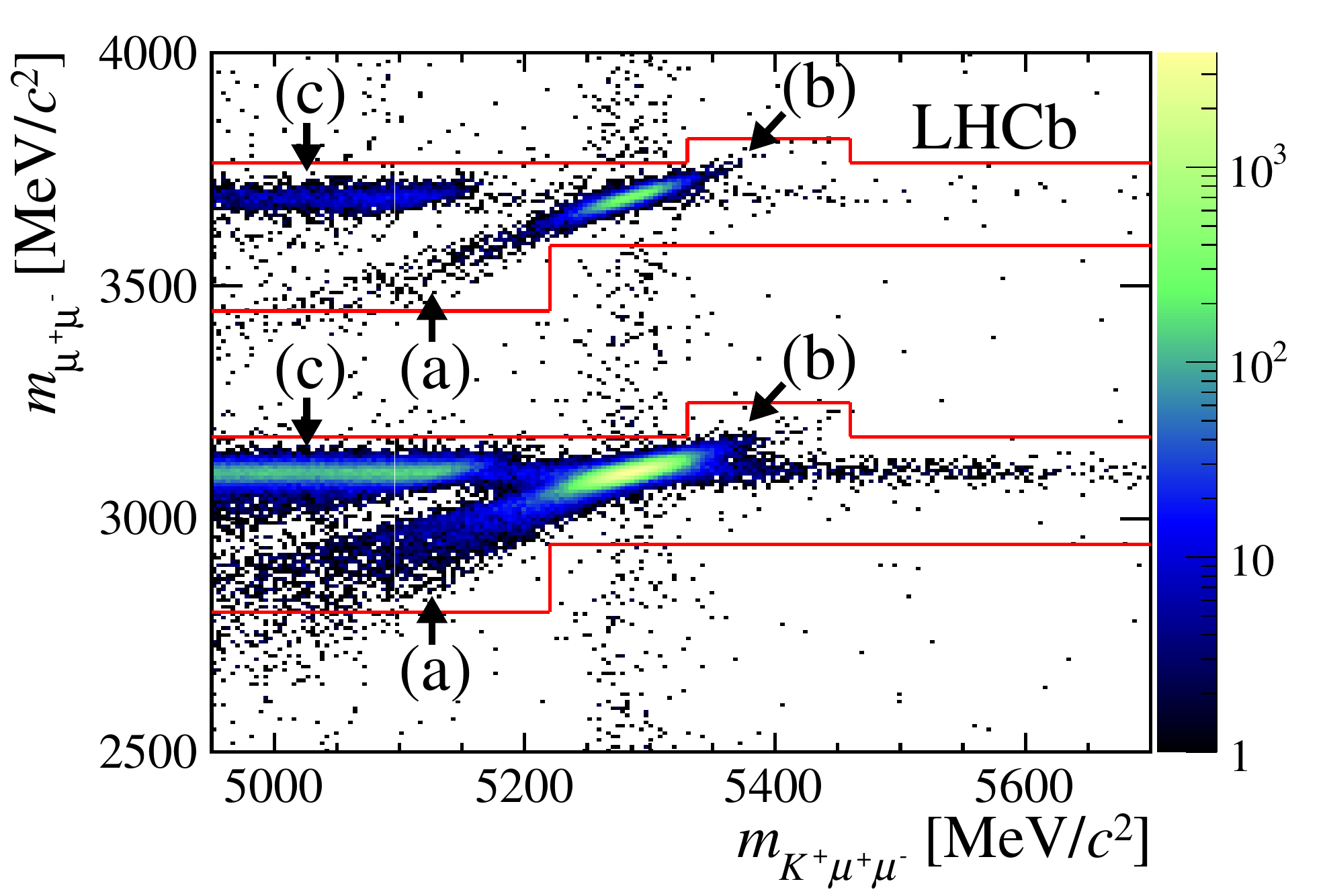}
\caption{Mass of the di-muon versus the mass of the \BuKMuMu
  candidates. Only the di-muon mass region close to the \jpsi and
  \psitwos masses is shown. The lines show the boundaries of the
  regions which are removed. Regions (a)--(c) are explained in the
  text.}
  \label{fig:2d}
\end{figure}
 
The mass distribution of \B candidates is shown versus the di-muon mass for \BuKMuMu 
data in Fig.~\ref{fig:2d}. The other signal channels have similar distributions, but with a smaller number of events.
The excess of candidates seen as horizontal bands around 3090\mevcc and 3690\mevcc are due to 
\jpsi and \psitwos decays, respectively. These events are removed from the signal channels by excluding the di-muon regions in the ranges 
2946$-$3181\mevcc and 3586$-$3766\mevcc. If a \B candidate has an mass below 5220\mevcc
 the veto is extended to 2800$-$3181\mevcc and 3450$-$3766\mevcc to eliminate candidates 
 for which the \jpsi or the \psitwos decay undergoes final state radiation. Such events are shown in Fig.~\ref{fig:2d} as regions (a).
In a small fraction of events, the di-muon mass is poorly 
reconstructed. This causes the \jpsi and \psitwos decay to leak into the 
region just above the \B mass. These events are shown in Fig.~\ref{fig:2d} as regions (b). 
The veto is extended to 2946$-$3250\mevcc and 3586$-$3816\mevcc in the candidate \B mass region from 
5330$-$5460\mevcc to eliminate these events. These vetoes largely remove the charmonium resonances 
and reduce the combinatorial background. Regions (c) in Fig.~\ref{fig:2d} are composed of \mbox{$\B \to \jpsi \Kp X$} and \mbox{$\B \to \psitwos \Kp X$} decays where $X$
 is not reconstructed. In the subsequent analysis only candidates with masses above 5170\mevcc are included to avoid dependence on
the shape of this background.

%% file: massfits.tex

\section{Signal yield determination}
\label{sec:Fits}

\begin{table}[t]
\newcommand\Tstrut{\rule{0pt}{2.5ex}}
\newcommand\TTstrut{\rule{0pt}{3.2ex}}
\newcommand\Bstrut{\rule[-1.2ex]{0pt}{0pt}}
\newcommand\BBstrut{\rule[-1.8ex]{0pt}{0pt}}

  \centering
   \caption{Signal yields of the \AllModes decays. The upper bound of the highest \qq bin, $\qq_{\mathrm{max}}$, is 19.3\gevgevcccc and 23.0\gevgevcccc for \BKstMuMu and \BKMuMu, respectively. }
    \begin{tabular}{@{}c c r c  r c  c@{}  }
      \toprule 
      \qq range   & \multicolumn{2}{c}{\KsMuMu} & \KMuMu & \multicolumn{2}{c}\bukst & \KstMuMu\Tstrut \\ 
      	 $[\gevgevcccc]$ \Tstrut	 & L  & \multicolumn{1}{c}D & L + D & \multicolumn{1}{c}L & D & L + D		\\ \midrule
$\phantom{0}0.05 - \phantom{0}2.00$ & $\phantom{1}1\pm2$ & $\phantom{1}2\pm3\phantom{1}$ & $\phantom{1}135\pm13$ &$4\pm3$ & $\phantom{1}5\pm4\phantom{1}$ & $108\pm11$  \\
$\phantom{0}2.00 - \phantom{0}4.30$ & $\phantom{1}2\pm3$ & $-1\pm3\phantom{1}$ & $\phantom{1}175\pm16$ & $3\pm2$ & $\phantom{1}5\pm3\phantom{1}$ &   \phantom{0}$53\pm \phantom{1}9$\\
$\phantom{0}4.30 - \phantom{0}8.68$ &  $\phantom{1}9\pm4$ & $16\pm6\phantom{1}$ & $\phantom{1}303\pm22$ & $4\pm3$ & $17\pm6\phantom{1}$ &  $203\pm17$\\
$10.09 - 12.86$ & $\phantom{1}4\pm3$ & $10\pm4\phantom{1}$ &  $\phantom{1}214\pm18$ & $4\pm3$ & $15\pm5\phantom{1}$ & $128\pm14$ \\
$14.18 - 16.00$ & $\phantom{1}3\pm2$ & $\phantom{1}3\pm3\phantom{1}$ & $\phantom{1}166\pm15$ & $5\pm3$ & $\phantom{1}4\pm3\phantom{1}$ &   \phantom{0}$90\pm10$\\
$16.00 - \phantom{1}\qq_{\mathrm{max}}$ & $\phantom{1}5\pm3$ & $\phantom{1}4\pm3\phantom{1}$ & $\phantom{1}257\pm19$ & $2\pm1$ & $\phantom{1}4\pm3\phantom{1}$ & \phantom{0}$80\pm11$ \\ \midrule
$\phantom{0}1.00 - \phantom{0}6.00$ & $\phantom{1}8\pm4$ & $\phantom{1}3\pm6\phantom{1}$ & $\phantom{1}356\pm23$ & $5\pm3$ & $15\pm5\phantom{1}$ &  $155\pm15$\\ \midrule
$\phantom{0}0.05 - \phantom{1}\qq_{\mathrm{max}}$ & $25\pm8$ & $35\pm11$ & $1250\pm42$ & $23\pm6$ & $53\pm10$ &  $673\pm30$\\
      \bottomrule
    \end{tabular}
  \label{tab:yields}
\end{table}

\begin{figure}[t] 
 \centering 
 \subfigure{\includegraphics[width=0.48\textwidth]{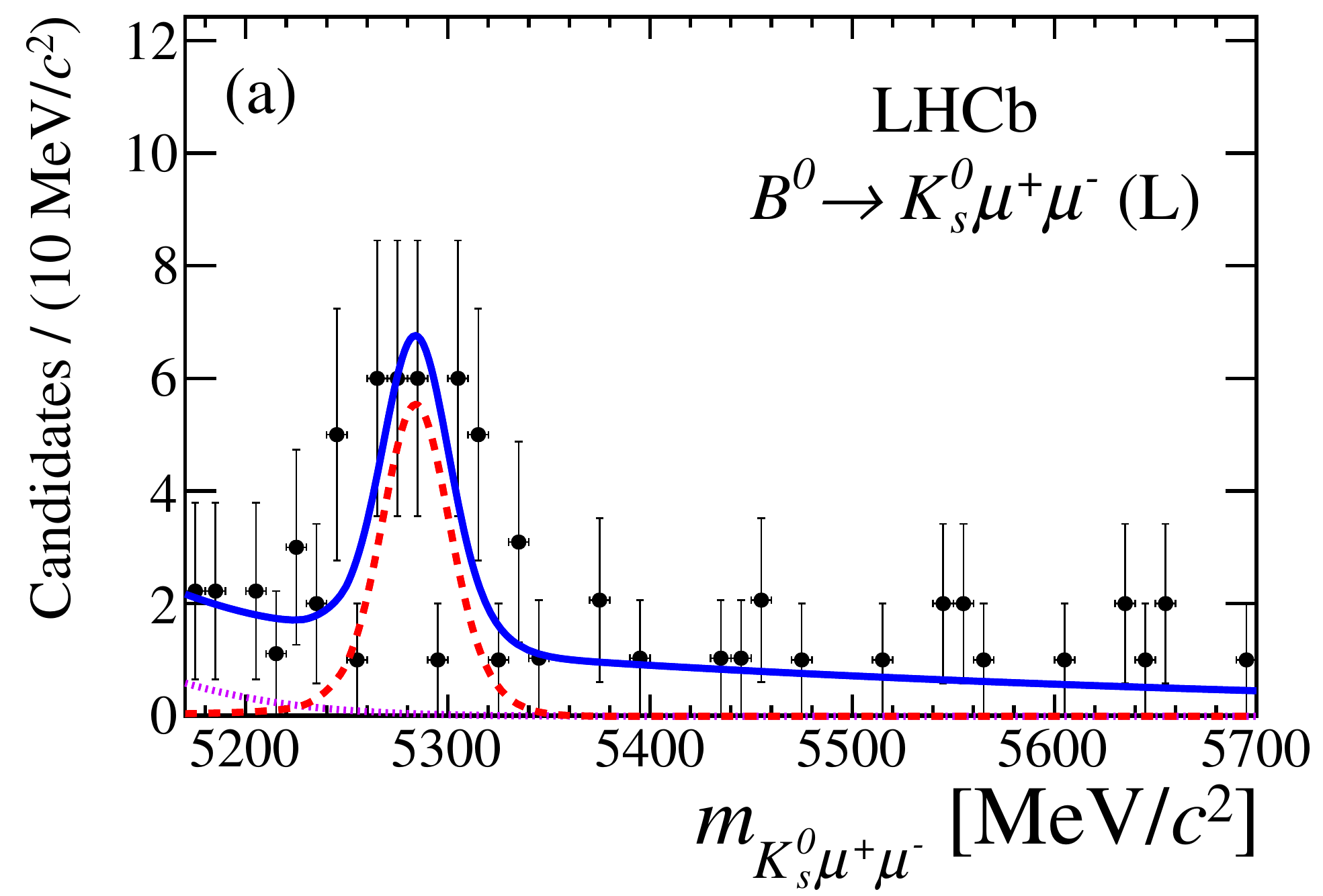}
  \label{fig:ksLL}} 
 \subfigure{\includegraphics[width=0.48\textwidth]{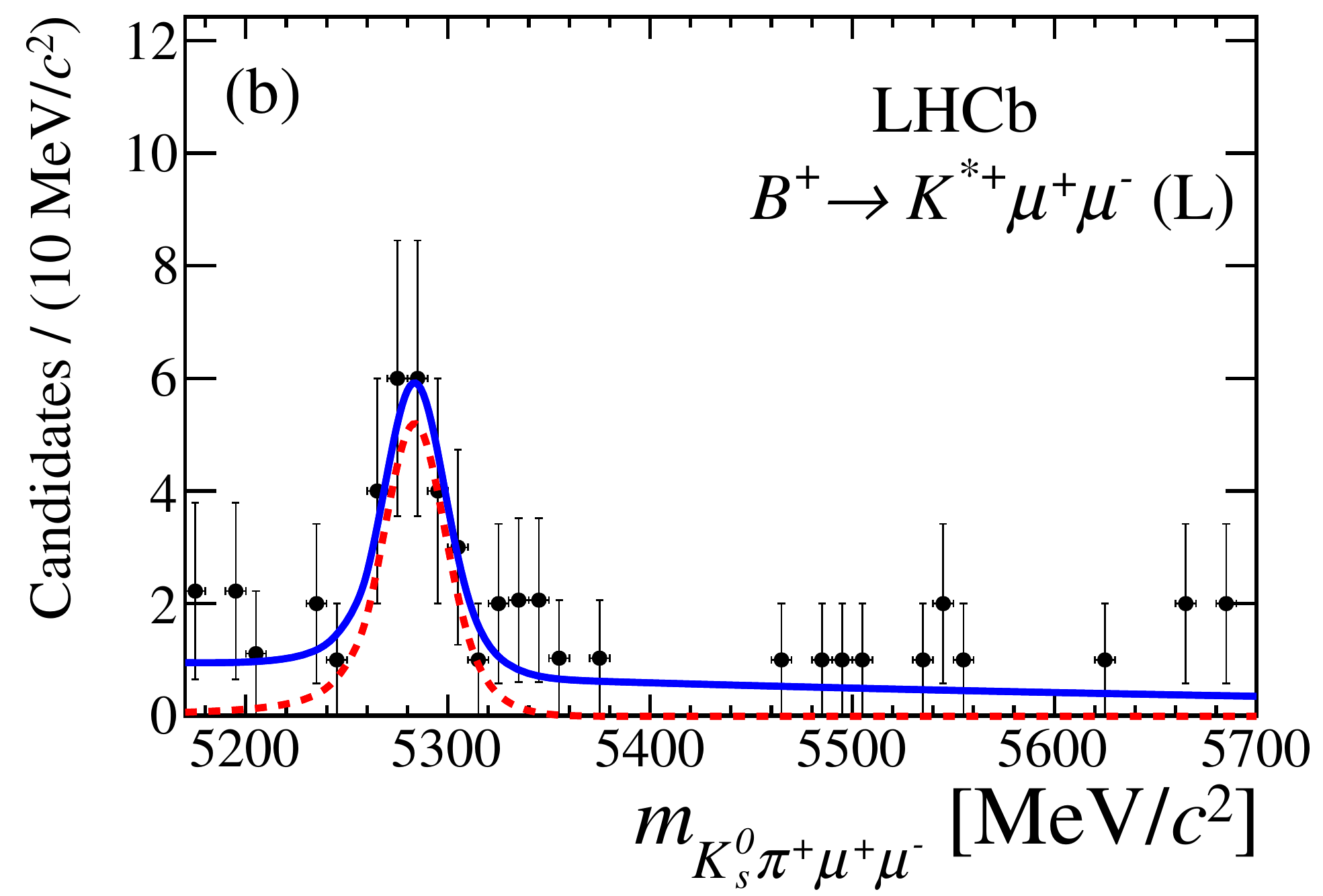}} \\
  \subfigure{\includegraphics[width=0.48\textwidth]{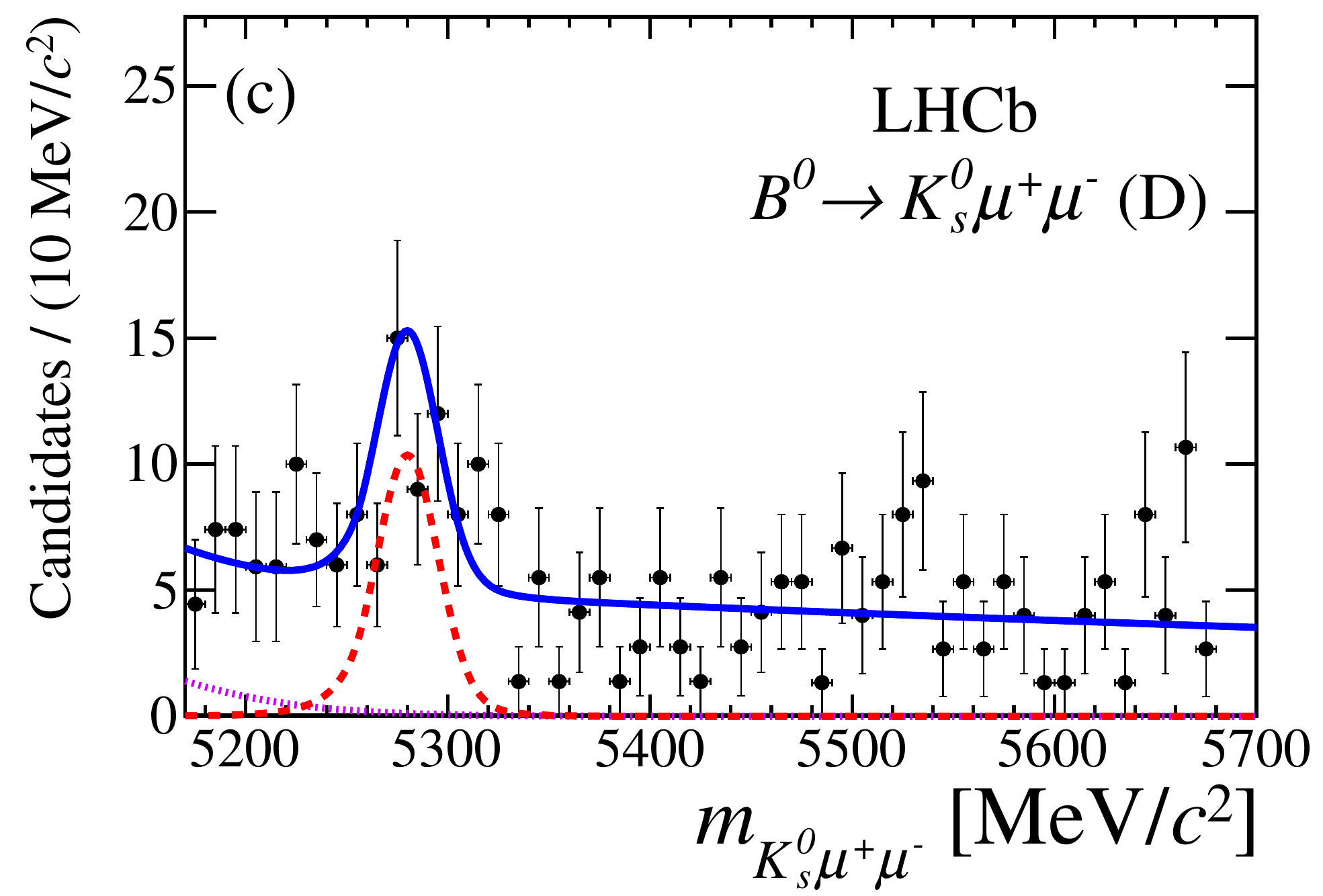}
 \label{fig:ksDD}}
 \subfigure{\includegraphics[width=0.48\textwidth]{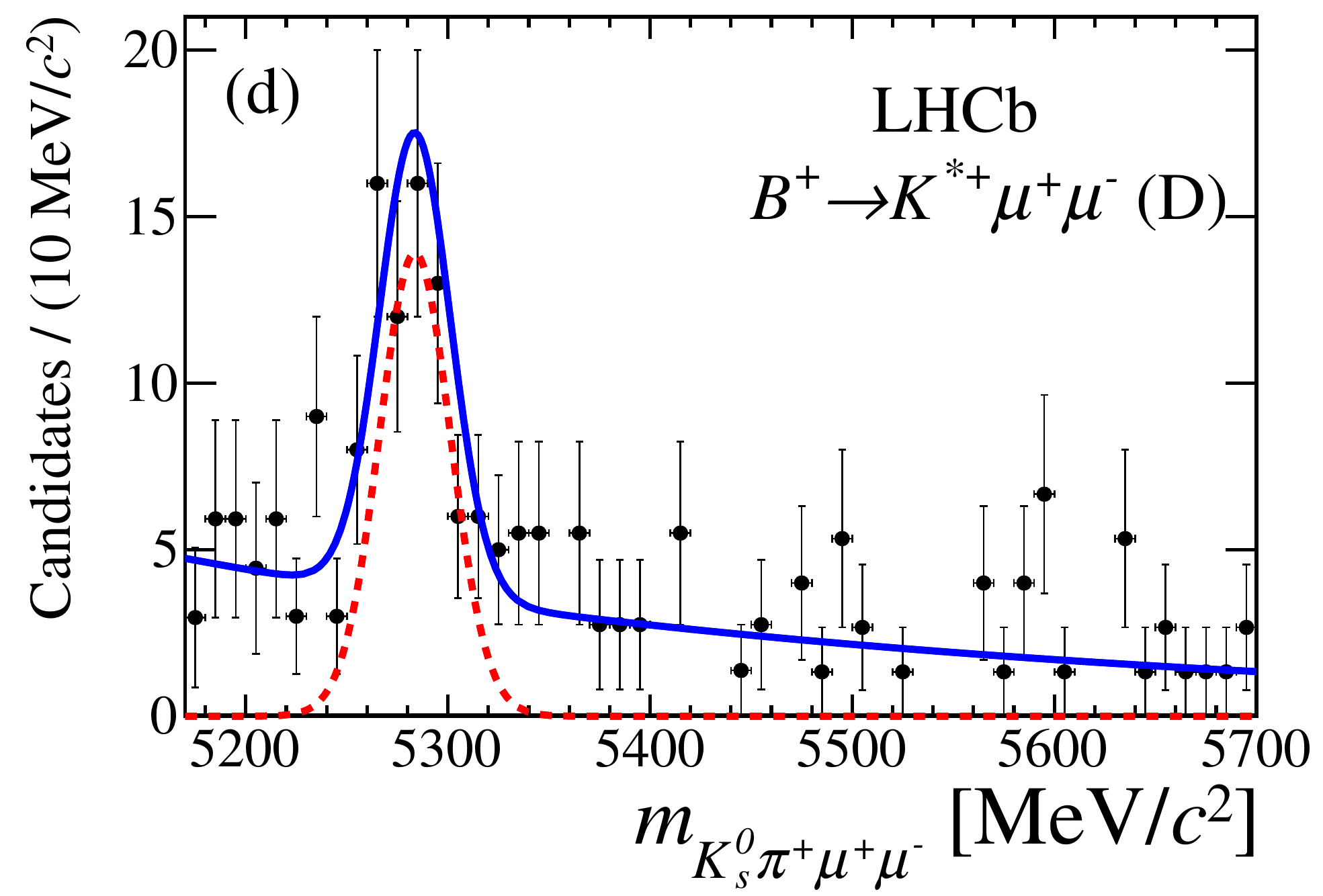}} \\
  \subfigure{\includegraphics[width=0.48\textwidth]{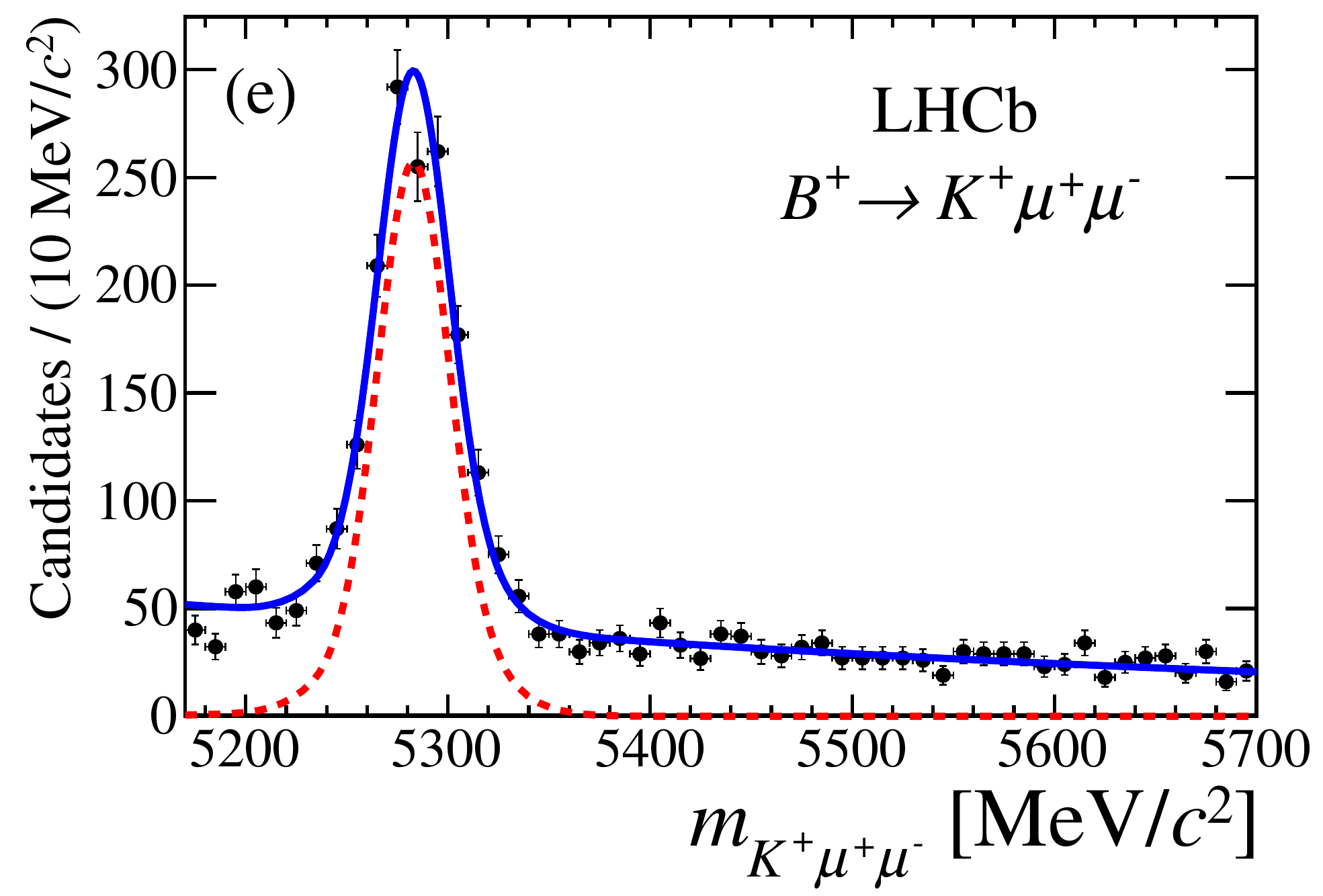}}
 \subfigure{\includegraphics[width=0.48\textwidth]{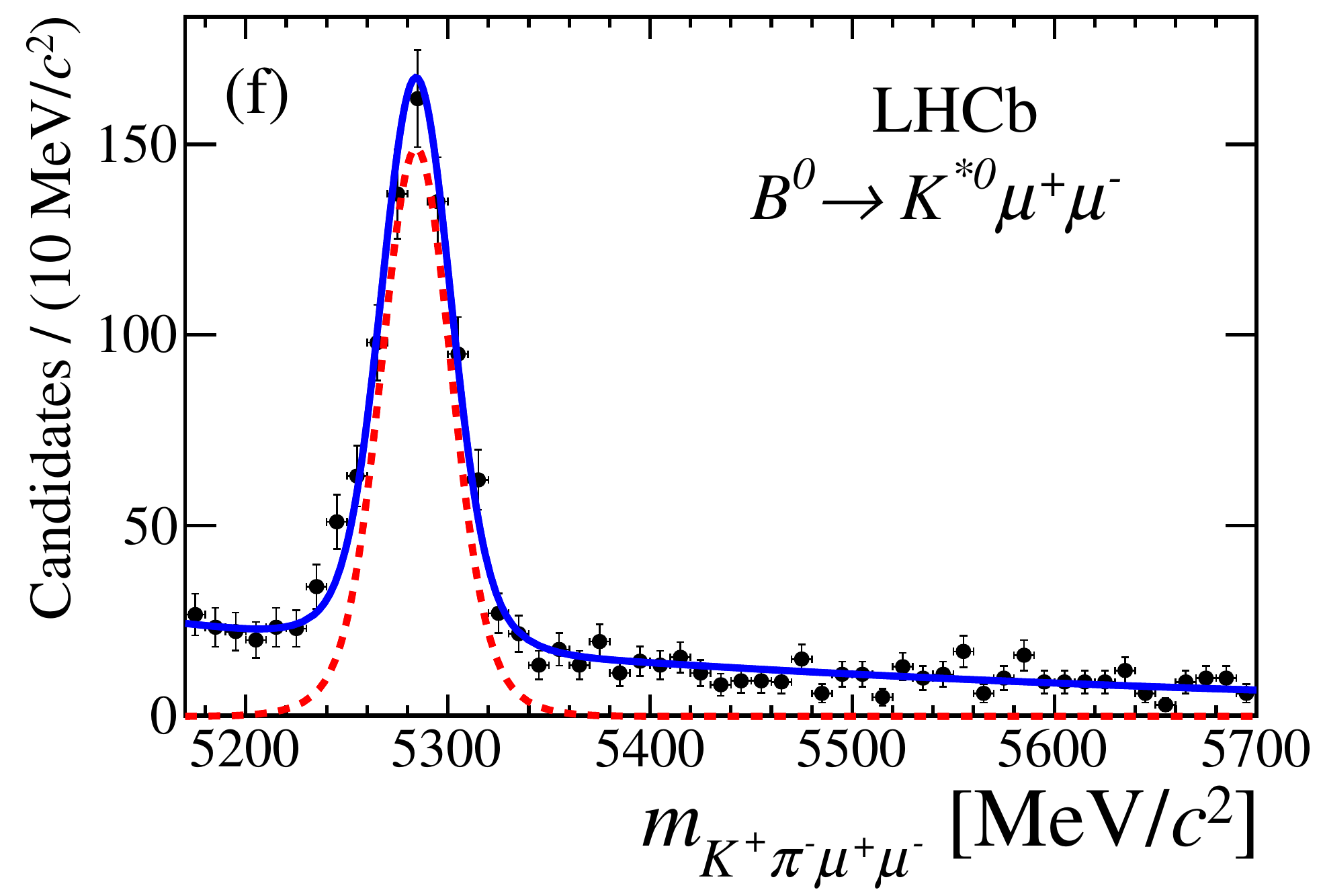}} \\
 \caption{Mass distributions and fits of the signal channels
   integrated over the full \qq region. For the \KS channels, the
   plots are shown separately for the L and D \KS reconstruction
   categories, (a,b) and (c,d) respectively. The signal component is
   shown by the dashed line, the partially reconstructed component
   in~\ref{fig:ksLL} and~\ref{fig:ksDD} is shown by the dotted line
   while the solid line shows the entire fit model.}
  \label{fig:mass} 
\end{figure}

The yields for the signal channels are determined using extended
unbinned maximum likelihood fits to the $\kaon^{(*)}\mumu$ mass in the
range 5170--5700\mevcc. These fits are performed in seven \qq bins and
over the full range as shown in Table~\ref{tab:yields}. The results of
the fits integrated over the full \qq range are shown in
Fig.~\ref{fig:mass}. After selection, the mass of \KS candidates is constrained
to the nominal \KS mass.  The signal component is described by the sum
of two Crystal Ball functions \cite{Skwarnicki:1986xj} with common
peak and tail parameters, but different widths.  The shape is taken to
be the same as the \mbox{\JpsiModes} normalisation channels. The combinatorial
background is fitted with a single exponential function. As stated in
Sect.~\ref{sec:Selection}, part of the combinatorial background is
removed by the charmonium vetoes.  This is accounted for by scaling
the remaining background. For the \mbox{\BKMuMu} decays, a component
arising mainly from partially reconstructed \BKstMuMu decays is
present at masses below the \B mass. This partially reconstructed
background is characterised using a threshold model detailed in
Ref.~\cite{patrick_kopp}. The shape of the partial reconstruction
component is again assumed to be the same as for the normalisation
channels. For the \BuKMuMu channel, the impact of this component is
negligible due to the relatively high signal and low background
yields. For the \BdKsMuMu channel, the amount of partially
reconstructed decays is found to be less than $25\%$ of the total
combinatorial background in the fit range.

The signal-shape parameters are allowed to vary in the \BdJpsiKs mass
fits and are subsequently fixed for the \BdKsMuMu mass fits when
calculating the significance. The significance $\sigma$ of a signal
$S$ for \BdKsMuMu is defined as \mbox{$\sigma^{2} =
  2\mathrm{ln}\mathcal{L}^{\textrm{L}}(S) +
  2\mathrm{ln}\mathcal{L}^{\textrm{D}}(S)-2\mathrm{ln}\mathcal{L}^{\textrm{L}}(0)
  - 2\mathrm{ln}\mathcal{L}^{\textrm{D}}(0)$} where
$\mathcal{L}^{\textrm{L,D}}(S)$ and $\mathcal{L}^{\textrm{L,D}}(0)$
are the likelihoods of the fit with and without the signal component,
respectively. The \mbox{\BdKsMuMu} channel is observed with a
significance of 5.7\,$\sigma$.

%% file: normalisation.tex

\section{Normalisation}
\label{sec:Normalisation}

In order to simplify the calculation of systematic uncertainties, each signal mode is normalised to the \JpsiModes channel, where the \jpsi decays into two muons. These decays have well measured branching fractions which are approximately two orders of magnitude higher than those of the signal decays. Each normalisation channel has similar kinematics and the same final state particles as the signal modes.

\begin{figure}[ht] 
 \centering 
 \subfigure{\includegraphics[width=0.49\textwidth]{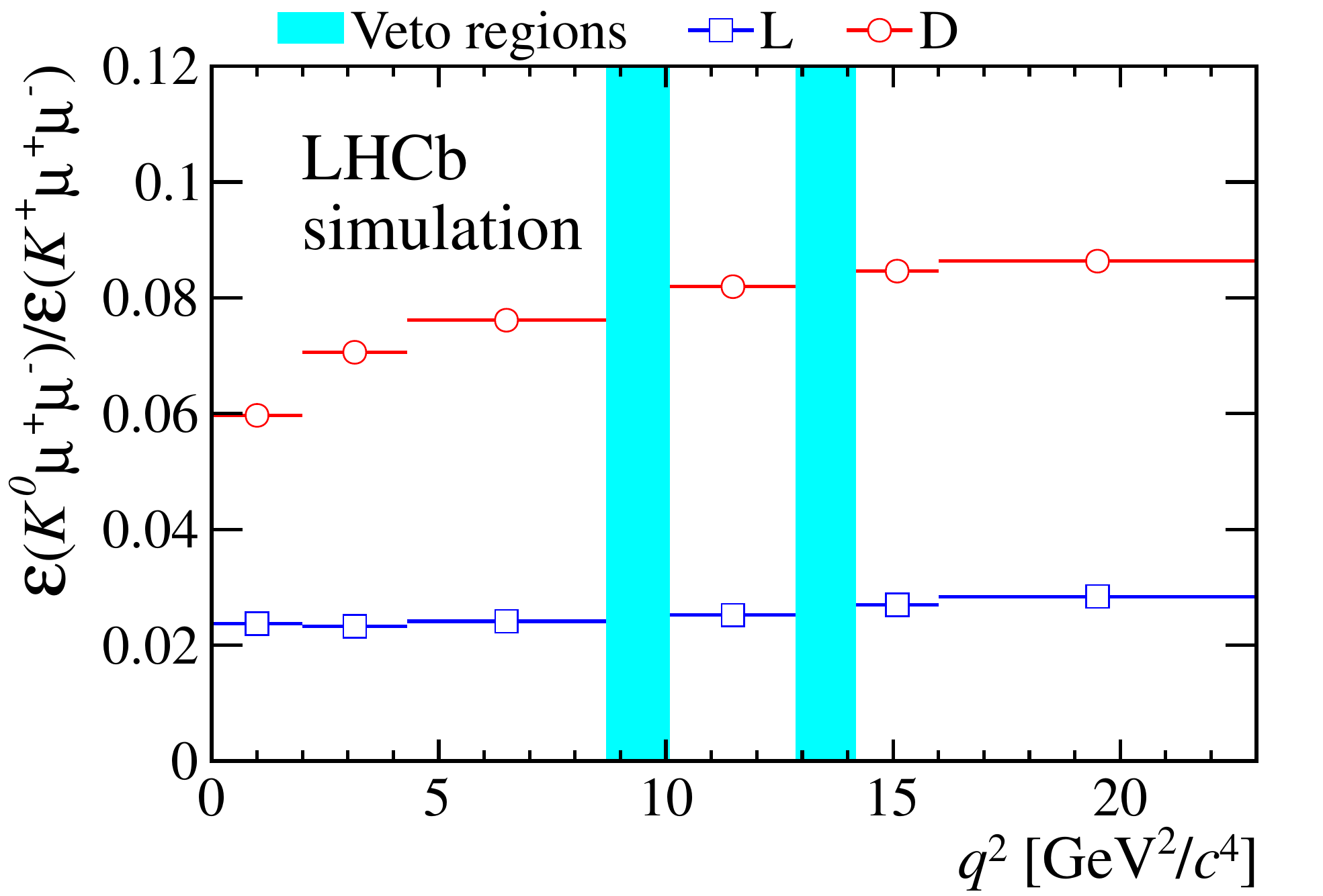}} 
 \subfigure{\includegraphics[width=0.49\textwidth]{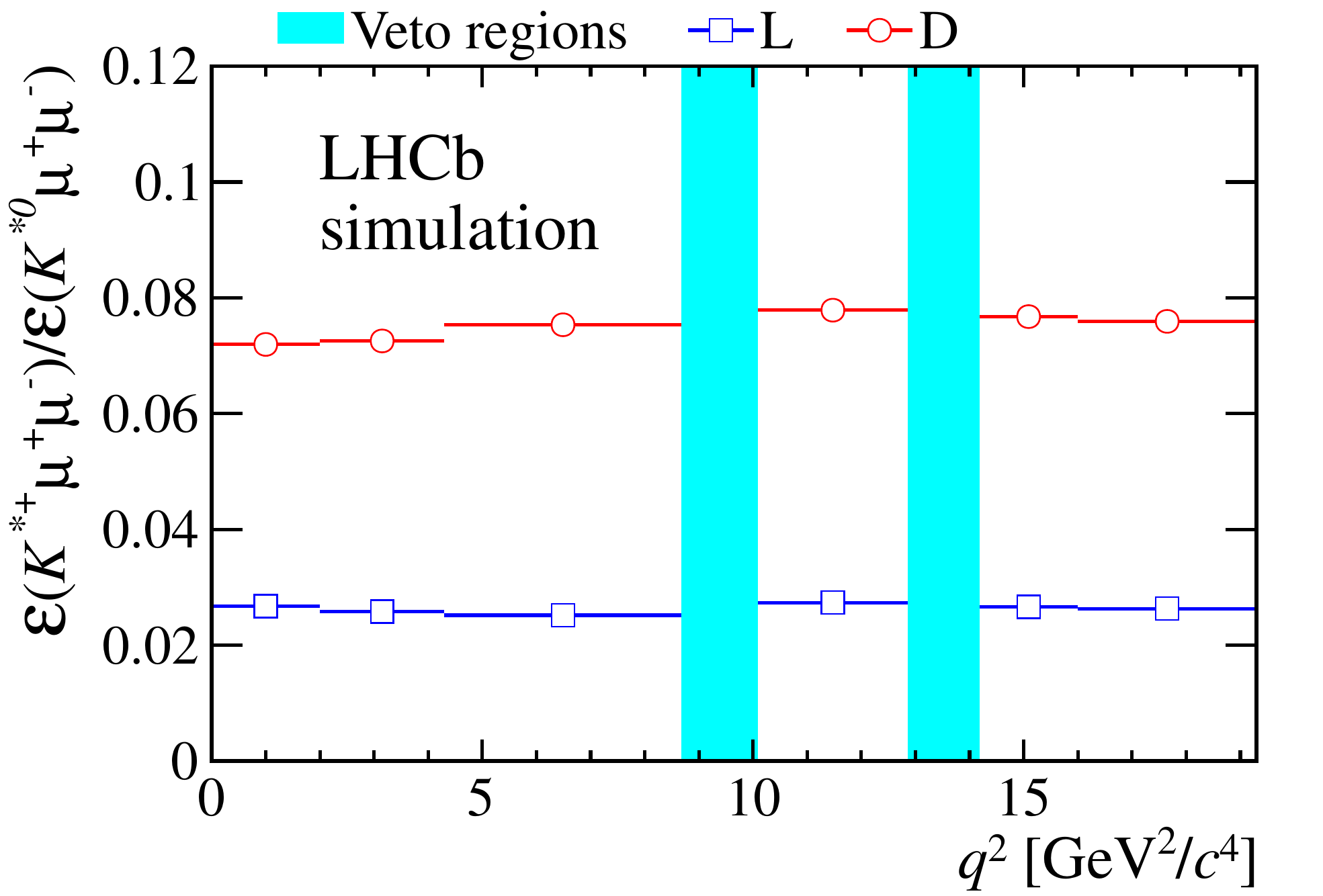}} \\
  \caption{Efficiency of the \KS channels with respect to the \Kp channels for (left) \mbox{\BKMuMu} and (right) \mbox{\BKstMuMu}, calculated using the simulation. The efficiencies are shown for both L and D \KS reconstruction categories and include the visible branching fraction of $\Kz\to\KS\to\pip\pim$. The error bars are not visible as they are smaller than the marker size.}
  \label{fig:eff} 
\end{figure}

The relative efficiency between signal and normalisation channels is
estimated using simulated events. After smearing the IP resolution of
all tracks by 20\%, the IP distributions of candidates in the
simulation and data agree well. The performance of the PID is studied
using the decay \decay{\Dstarp}{(\decay{\Dz}{\pip \Km}) \pip}, which
provides a clean source of kaons to study the kaon PID efficiency, and
a \emph{tag-and-probe} sample of \BuJpsiK to study the muon PID
efficiency. The simulation is reweighted to match the PID performance
of the data.

Integrating over \qq, the relative efficiency between the signal and
normalisation channels is between 70 and 80\% depending on the decay
mode and category. The relative efficiency includes differences in the
geometrical acceptance, as well as the reconstruction, selection and
trigger efficiencies. Most of these effects cancel in the efficiency
ratio between \KS and \Kp channels, as shown in
Fig.~\ref{fig:eff}. The dominant effect remaining is due to the \KS
reconstruction efficiency, which depends on the \KS momentum. At low
\qq, the efficiency for \BdKsMuMu (D) decreases with respect to that
for \BuKMuMu due to the high \KS momentum in this region. This results
in the \KS meson more often decaying beyond the TT and consequently it
has a lower reconstruction efficiency. This effect is not seen in the
\BuKstMuMu D category as the \KS typically has lower momentum in this
decay and so the \KS reconstruction efficiency is approximately
constant across \qq. This \KS reconstruction effect is also seen in
the L category for both modes but is partially compensated by the fact
that the \KS daughters can cause the event to be triggered, which
increases the trigger efficiency with respect to the \Kp channels at
low \qq. Summed over both the L and D categories, the efficiency of
the decays involving a \Kz meson is approximately 10\% with respect to
those involving a charged kaon. This is partly due to the visible
branching fraction of $\Kz\to\KS\to\pip\pim$ ($\sim$30\%) and partly
due to the lower reconstruction efficiency of the \KS due to the long
lifetime and the need to reconstruct an additional track ($\sim$30\%).
The relative efficiency between the L and D signal categories is
cross-checked by comparing the ratio for the \decay{\B}{\psitwos
  K^{(*)}} decay to the corresponding ratio for the \JpsiModes decays
seen in data. The results agree within the statistical accuracy of
5\%.

%% file: systematics.tex

\section{Systematic uncertainties}
\label{sec:Systematics}

Gaussian constraints are used to include all systematic uncertainties in the fits for \AI and the 
branching fractions. In most 
cases the dominant systematic uncertainty is that from 
the branching fraction measurements of the normalisation channels, 
ranging from 3 to 6\%. There is also a 
statistical uncertainty on the yield of the normalisation channels, 
which is in the range 0.5--2.0\%, depending on the channel.

The finite size of the simulation samples introduces a 
statistical uncertainty on the relative efficiency and leads to a 
systematic uncertainty in the range 0.8--2.5\% depending on \qq and decay mode.

The relative tracking efficiency between the signal and normalisation 
channels is corrected using data. The statistical precision of these 
corrections leads to a systematic uncertainty of $\sim$ 0.2\% per 
long track. The differences between the downstream tracking efficiency 
between the simulation and data are expected to mostly cancel in the normalisation procedure. A conservative systematic uncertainty of 1\% per 
downstream track is assigned for the variation across \qq.

The PID efficiency is derived from data, and its corresponding systematic uncertainty
 arises from the statistical error associated with the PID 
efficiency measurements. 
The uncertainty on the relative efficiency is determined by randomly varying 
PID efficiencies within their uncertainties, and recomputing the relative efficiency. 
The resulting uncertainty is found to be negligible.

The trigger efficiency is calculated using the simulation. Its uncertainty consists of two components, 
one associated with the trigger efficiency of the \KS meson, and one associated 
with the trigger efficiency of the muons (and pion from the \Kstar). 
For the muons and pion the uncertainty is obtained using \BuJpsiK 
and \BdJpsiKst events in data that are triggered independently of the signal. 
These candidates are used to calculate the trigger efficiency and are 
compared to the efficiency 
calculated using the same method in simulation. The difference is found to be $\sim2\%$ for both 
\BuJpsiK and \BdJpsiKst decays and is 
assigned as a systematic uncertainty. This uncertainty is assumed to cancel for 
the isospin asymmetry as the presence of muons is common between 
the \KS channels and the \Kp channels. The uncertainty associated with the \KS trigger 
efficiency is calculated by comparing the fraction of candidates triggered by \KS daughters 
in the simulation and the data. The difference is 
used as an estimate of the capability of simulation to reproduce these trigger 
decisions. The simulation is found to underestimate the \KS trigger 
decisions by 10--20\% depending on the decay mode. This percentage 
is multiplied by the fraction of trigger decisions where the \KS 
participates in a given bin of \qq leading to an uncertainty of 0.2--4.1\% 
depending on \qq and decay mode.

The effect of the unknown angular distribution of \BuKstMuMu decays on the relative 
efficiency is estimated by altering the Wilson coefficients 
appearing in the operator product expansion method~\cite{wilson,wilson2}. 
The Wilson coefficients, $\mathcal{C}_{7}$ 
and $\mathcal{C}_{10}$, have their real part inverted and the relative efficiency is recalculated. 
This can be seen as an extreme variation which is used to obtain a conservative estimate of the associated uncertainty.
 The calculation was performed using an \evtgen physics model which uses the transition form 
  factors detailed in Ref.~\cite{nikolai}. The difference in the relative efficiency
   varies from 0--6\%, depending on \qq, and it is assigned as a systematic uncertainty.  

The shape parameters for the signal modes are assumed to be the same as the 
normalisation channels. This assumption is validated using the simulation 
and no systematic uncertainty is assigned. The statistical uncertainties of these shape 
parameters are propagated through the fit using Gaussian constraints, 
accounting for correlations between the parameters. The uncertainty 
on the amount of partially reconstructed background is also added 
to the fit using Gaussian constraints, therefore no further uncertainty is added. 
The parametrisation of the fit model is cross-checked by varying the fit 
range and background model. Consistent yields are observed and 
no systematic uncertainty is assigned.

Overall the systematic error on the branching fraction is 4--8\% depending on \qq and the decay mode. 
This is small compared to the typical statistical error of $\sim$ 40\%.

%% file: results.tex

\section{Results and conclusions}
\label{sec:Results}

\begin{figure}[t] 
 \centering 
 \subfigure{\includegraphics[width=0.49\textwidth]{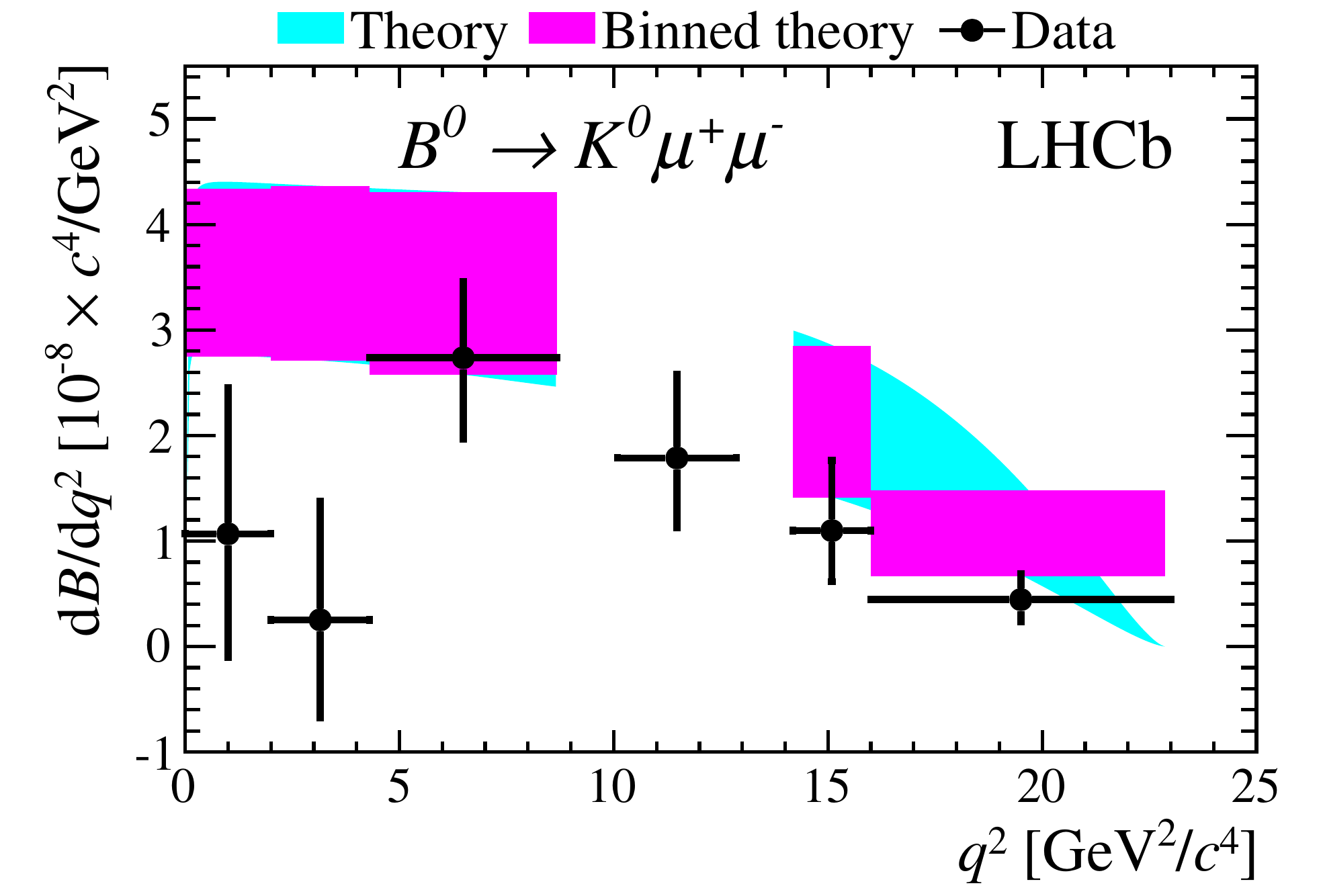}} 
 \subfigure{\includegraphics[width=0.49\textwidth]{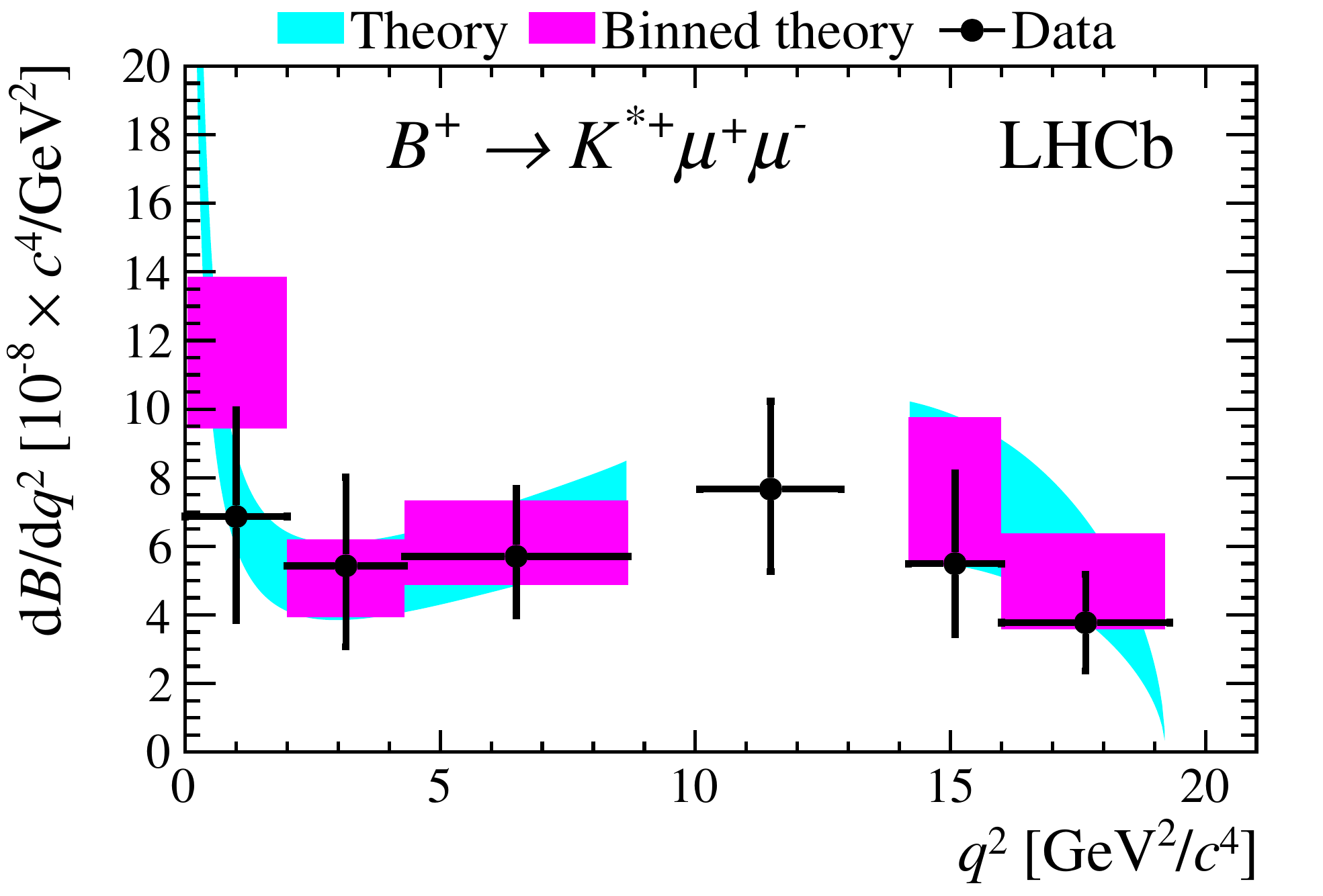}} \\
  \caption{Differential branching fractions of (left) \mbox{\BdKzMuMu} and (right) \mbox{\BuKstMuMu}. The theoretical SM predictions are taken from Refs.~\cite{Bobeth:2011gi,Bobeth:2011nj}.}
  \label{fig:BF} 
\end{figure}

\begin{figure}[t] 
 \centering 
 \subfigure{\includegraphics[width=0.49\textwidth]{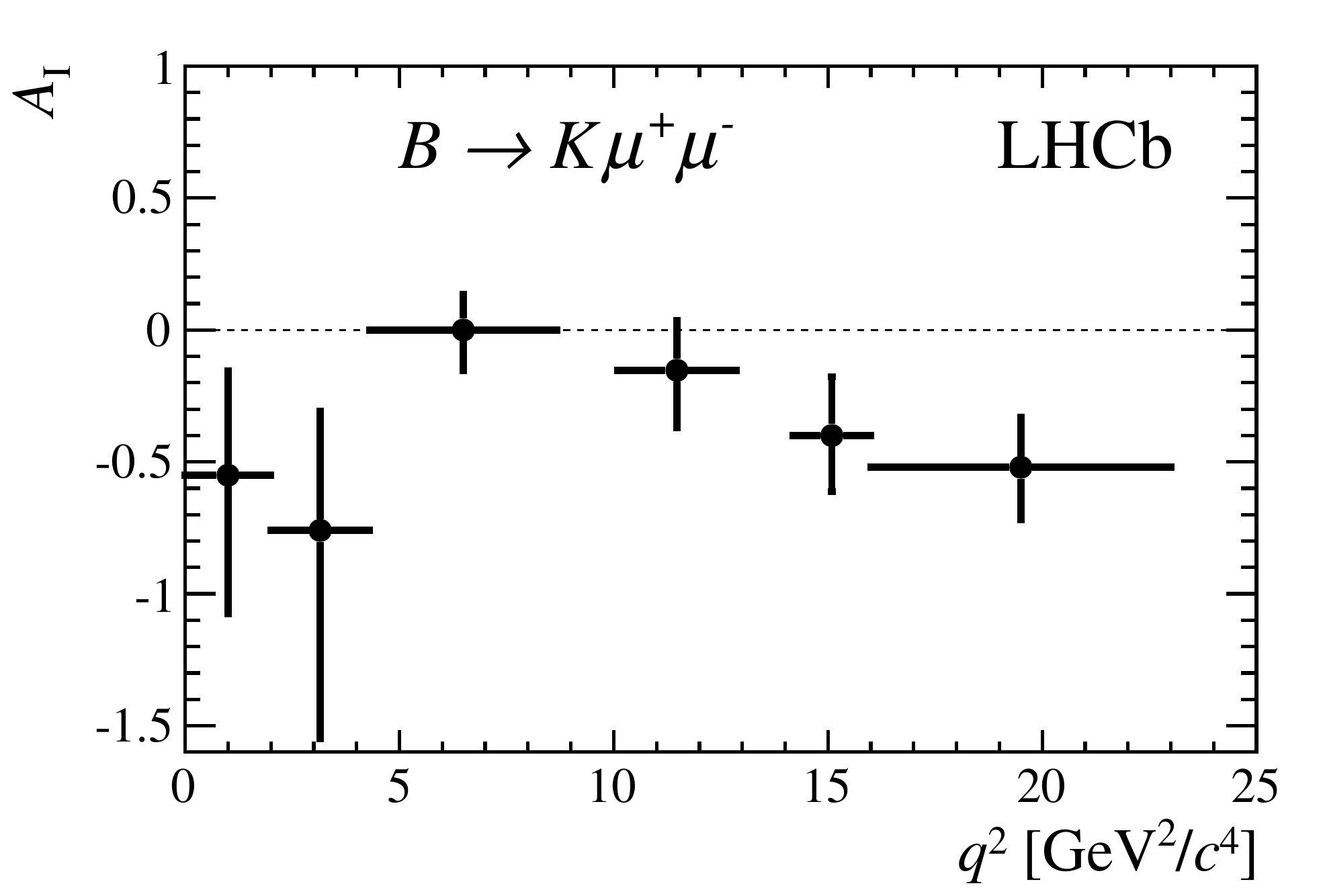}} 
 \subfigure{\includegraphics[width=0.49\textwidth]{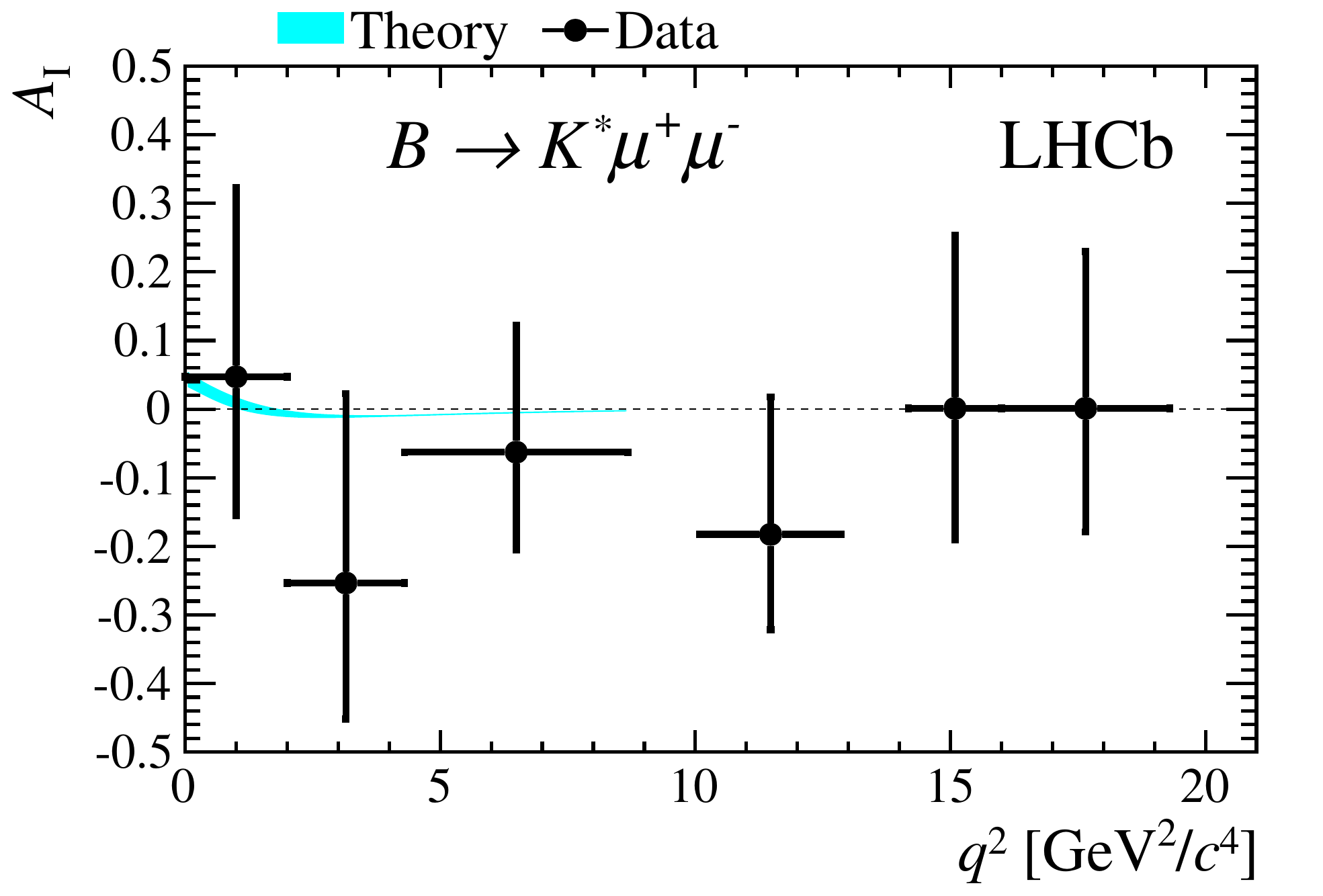}} \\
  \caption{Isospin asymmetry of (left) \mbox{\BKMuMu} and (right) \mbox{\BKstMuMu}. For \mbox{\BKstMuMu} the theoretical SM prediction, which is very close to zero, is shown for \qq below 8.68\gevcc, from Ref.~\cite{theorists}.}
  \label{fig:AI} 
\end{figure}

The differential branching fraction in the $i^{\mathrm{th}}$ \qq bin can be written as

\begin{equation} 
\frac{d\mathcal{B}^{i}}{dq^{2}} = \frac{N^{i}(\AllModes)}{N(\JpsiModes)}\times\frac{\mathcal{B}(\JpsiModes)\mathcal{B}(\jpsi\to\mup\mun)}{\epsilon^{i}_{\mathrm{rel}}\Delta^{i}},
\label{eq:BF}
\end{equation}
where ${N^{i}(\AllModes})$ is the number of signal candidates in bin
$i$, ${N(\JpsiModes})$ is the number of normalisation candidates, the
product of \mbox{$\mathcal{B}(\JpsiModes)$} and
\mbox{$\mathcal{B}(\jpsi\to\mup\mun)$} is the visible branching
fraction of the normalisation channel~\cite{Nakamura:2010zzi},
$\epsilon^{i}_{\mathrm{rel}}$ is the relative efficiency between the
signal and normalisation channels in bin $i$ and finally $\Delta^{i}$
is the bin $i$ width.  The differential branching fraction is
determined by simultaneously fitting the L and D categories of the
signal channels. The branching fraction of the signal channel is
introduced as a fit parameter by re-arranging Eq.~(\ref{eq:BF}) in
terms of ${N(\AllModes})$. Confidence intervals are evaluated by
scanning the profile likelihood. The results of these fits for
\BdKzMuMu and \BuKstMuMu decays are shown in Fig.~\ref{fig:BF} and
given in Tables~\ref{tab:kmumu} and~\ref{tab:kstmumu}.  Theoretical
predictions~\cite{Bobeth:2011gi,Bobeth:2011nj,theorists} are
superimposed on Figs.~\ref{fig:BF} and~\ref{fig:AI}. In the low \qsq
region, these predictions rely on the QCD factorisation approaches
from Refs.~\cite{Beneke:2001at,Beneke:2004dp} for \BKstMuMu and
Ref.~\cite{Bobeth:2007dw} for \BKMuMu which lose accuracy when
approaching the \jpsi resonance. In the high \qsq region, an operator
product expansion in the inverse $b$-quark mass, $1/m_\bquark$, and in
$1/\sqrt{\qsq}$ is used based on Ref.~\cite{Grinstein:2004vb}. This
expansion is only valid above the open charm threshold. In both \qq
regions the form factor calculations for \BKstMuMu and \BKMuMu are
taken from Refs.~\cite{Ball:2004rg} and~\cite{Khodjamirian:2010vf}
respectively. These form factors lead to a high correlation in the 
uncertainty of the predictions across \qq. A dimensional estimate is made of the uncertainty from
expansion corrections~\cite{Egede:2008uy}.  The non-zero isospin
asymmetry arises in the low \qq region due to spectator-quark
differences in the so-called hard-scattering part. There are also
sub-leading corrections included from Refs.~\cite{isospintheory}
and~\cite{Beneke:2004dp} which only affect the charged modes
and further contribute to the isospin asymmetry.

The total branching fractions are also measured by extrapolating
underneath the charmonium resonances assuming the same \qq
distribution as in the simulation. The branching fractions of
\BdKzMuMu and \BuKstMuMu are found to be

  \begin{equation}
  \begin{split}
      \phantom{+}  \mathcal{B}(\BdKzMuMu) = (0.31^{+0.07}_{-0.06}) \times 10^{-6} \phantom{,\pm b}  \rm{and} \\ 
	\mathcal{B}(\BuKstMuMu) = (1.16\pm0.19) \times 10^{-6}, \phantom{and} \nonumber
  \end{split}
    \end{equation}
    respectively, where the errors include statistical and systematic uncertainties. 
    These results are in agreement with previous measurements and with better precision~\cite{Nakamura:2010zzi}.

The isospin asymmetries as a function of \qq for \BKMuMu and \BKstMuMu
are shown in Fig.~\ref{fig:AI} and given in Tables~\ref{tab:kmumu}
and~\ref{tab:kstmumu}.  As for the branching fractions, the fit is
done simultaneously for both the L and D categories where \AI is a
common parameter for the two cases. The confidence intervals are also
determined by scanning the profile likelihood.  The significance of
the deviation from the null hypothesis is obtained by fixing \AI to be
zero and computing the difference in the negative log-likelihood from
the nominal fit.

In summary, the isospin asymmetries of \AllModes decays and the
branching fractions of \BdKzMuMu and \BuKstMuMu are measured, using
1.0\invfb of data taken with the LHCb detector.  The two \qq bins
below 4.3\gevcc and the highest bin above 16\gevcc have the most negative isospin
asymmetry in the \BKMuMu channel. These \qq regions are
furthest from the charmonium regions and are therefore cleanly
predicted theoretically. This asymmetry is dominated by a deficit in
the observed \BdKzMuMu signal. Ignoring the small correlation of
errors between each \qq bin, the significance of the deviation from
zero integrated across \qq is calculated to be 4.4 $\sigma$. The
\BKstMuMu case agrees with the SM prediction of almost zero isospin
asymmetry~\cite{isospintheory}. All results agree with previous
measurements~\cite{belleiso,babarnew,Aaltonen:2011qs}.

\begin{table}[t]

\newcommand\Tstrut{\rule{0pt}{2.5ex}}
\newcommand\TTstrut{\rule{0pt}{3.2ex}}
\newcommand\Bstrut{\rule[-1.2ex]{0pt}{0pt}}
\newcommand\BBstrut{\rule[-1.8ex]{0pt}{0pt}}

  \centering
   \caption{Partial branching fractions of \mbox{\BdKzMuMu} and isospin asymmetries of \mbox{\BKMuMu} decays. The significance of the deviation of \AI from zero is shown in the last column. The errors include the statistical and systematic uncertainties.}
    \begin{tabular}{ c  c  c  c  }
      \toprule 
      \qq range [\gevgevcccc\!]  & $d\mathcal{B}/d\qq [10^{-8}/\gevgevcccc]$ & \AI & $\sigma$(\AI = 0) \\  \midrule
$\phantom{0}0.05 - \phantom{0}2.00$ & $1.1^{+1.4}_{-1.2}$ & $-0.55^{+0.40}_{-0.56}$ & 1.5  \\
$\phantom{0}2.00 - \phantom{0}4.30$ & $0.3^{+1.1}_{-0.9}$\Tstrut & $-0.76^{+0.45}_{-0.79}$ & 1.9 \\
$\phantom{0}4.30 - \phantom{0}8.68$ & $2.8\pm0.7$\Tstrut & $\phantom{-}0.00^{+0.14}_{-0.15}$ & 0.1 \\
$10.09 - 12.86$ & $1.8^{+0.8}_{-0.7}$\Tstrut & $-0.15^{+0.19}_{-0.22}$ & 0.8 \\
$14.18 - 16.00$ & $1.1^{+0.7}_{-0.5}$\Tstrut & $-0.40\pm0.22$ & 1.9\\
$16.00 - 23.00$ & $0.5^{+0.3}_{-0.2}$\Tstrut & $-0.52^{+0.18}_{-0.22}$ & 3.0 \\ \midrule
$\phantom{0}1.00 - \phantom{0}6.00$ & $1.3^{+0.9}_{-0.7}$ & $-0.35^{+0.23}_{-0.27}$ & 1.7 \\
      \bottomrule
    \end{tabular}
  \label{tab:kmumu}
\end{table}

\begin{table}[t]
\newcommand\Tstrut{\rule{0pt}{2.5ex}}
\newcommand\TTstrut{\rule{0pt}{3.2ex}}
\newcommand\Bstrut{\rule[-1.2ex]{0pt}{0pt}}
\newcommand\BBstrut{\rule[-1.8ex]{0pt}{0pt}}
  \centering
  \caption{Partial branching fractions of \mbox{\BuKstMuMu} and isospin asymmetries of \mbox{\BKstMuMu} decays. The significance of the deviation of \AI from zero is shown in the last column. The errors include the statistical and systematic uncertainties.}
    \begin{tabular}{ c  c  c  c   }
      \toprule 
      \qq range [\gevgevcccc\!]  & $d\mathcal{B}/d\qq [10^{-8}/\gevgevcccc]$ & \AI & $\sigma$(\AI = 0)  \\ \midrule
$\phantom{0}0.05 - \phantom{0}2.00$ & $7.0^{+3.1}_{-3.0}$ & $\phantom{-}0.05^{+0.27}_{-0.21}$ & 0.2  \\
$\phantom{0}2.00 - \phantom{0}4.30$ & $5.4^{+2.6}_{-2.4}$\Tstrut & $-0.27^{+0.29}_{-0.18}$ & 0.9 \\
$\phantom{0}4.30- \phantom{0}8.68$ & $5.7^{+2.0}_{-1.7}$\Tstrut & $-0.06^{+0.19}_{-0.14}$ & 0.4 \\
$10.09 - 12.86$ & $7.7^{+2.6}_{-2.4}$ & $-0.16^{+0.17}_{-0.16}$\Tstrut & 0.9 \\
$14.18 - 16.00$ & $5.5^{+2.6}_{-2.1}$\Tstrut & $\phantom{-}0.02^{+0.23}_{-0.21}$ & 0.1 \\
$16.00 - 19.30$ & $3.8\pm1.4$\Tstrut & $\phantom{-}0.02^{+0.21}_{-0.20}$ & 0.1  \\ \midrule
$\phantom{0}1.00 - \phantom{0}6.00$ & $5.8^{+1.8}_{-1.7}$ & $-0.15\pm0.16$ & 1.0 \\
      \bottomrule
    \end{tabular}
  \label{tab:kstmumu}
\end{table}

%% file: acknowledgements.tex
\section*{Acknowledgements}

\noindent We would like to thank Christoph Bobeth, Danny van Dyk and Gudrun Hiller 
for providing SM predictions for the branching fractions and the isospin asymmetry of \BKstMuMu decays. We express 
our gratitude to our colleagues in the CERN accelerator
departments for the excellent performance of the LHC. We thank the
technical and administrative staff at CERN and at the LHCb institutes,
and acknowledge support from the National Agencies: CAPES, CNPq,
FAPERJ and FINEP (Brazil); CERN; NSFC (China); CNRS/IN2P3 (France);
BMBF, DFG, HGF and MPG (Germany); SFI (Ireland); INFN (Italy); FOM and
NWO (The Netherlands); SCSR (Poland); ANCS (Romania); MinES of Russia and
Rosatom (Russia); MICINN, XuntaGal and GENCAT (Spain); SNSF and SER
(Switzerland); NAS Ukraine (Ukraine); STFC (United Kingdom); NSF
(USA). We also acknowledge the support received from the ERC under FP7
and the Region Auvergne.